\documentclass[]{article}

\usepackage{dcolumn}
\usepackage{bm}
\usepackage{amssymb}
\usepackage{latexsym}
\usepackage{graphicx}

\title{\bf On the Geometry of the Space-Time \\ and Motion of the Spinning Bodies}

\author{Kostadin Tren\v{c}evski\footnote{E-mail: kostatre@pmf.ukim.mk} \\
Faculty of Natural Sciences and Mathematics, \\ Ss. Cyril and Methodius University,\\
P.O.Box 162, Skopje, Macedonia}

\date{}

\begin{document}

\maketitle

\begin{abstract}
In this paper an alternative theory about space-time is given. 
First some preliminaries about 3-dimensional time and the reasons for its introduction are presented. 
Alongside the 3-dimensional space (S) the 3-dimensional space of 
spatial rotations (SR) is considered independently from the 3-dimensional space.  
Then it is given a model of the universe, based on the Lie groups of real and complex 
orthogonal $3\times 3$ matrices in this 3+3+3-dimensional space.  
Special attention is dedicated for introduction and study of the space $S\times SR$, 
which appears to be isomorphic to $SO(3,{\mathbb R})\times SO(3,{\mathbb R})$ or $S^3\times S^3$. 
The influence of the gravitational acceleration to the spinning bodies is considered. 
Some important applications of these results about spinning 
bodies are given, which naturally lead to 
violation of Newton's third law in its classical formulation. 
The precession of the spinning axis is also considered. 


\noindent 2010MSC: 53A04, 53Z05, 51P05


\end{abstract}

\section{Introduction and motivation}
\label{sec:1}

The Special and the General Relativity give the geometry of the 3+1- dimensional space-time. 
The Special Relativity is free of any anomaly, if we assume that the space-time 
is homeomorphic to ${\mathbb R}^4$. The General Relativity appears to be very good when we 
consider trajectories of motion, gravitational radiation and so on. But if we consider 
the precession of axis of a gyroscope, we have a different situation. Using the 
Fermi-Walker connection, it is well known that the angular velocity of the axis of a gyroscope 
in a free fall orbit is given by 
(\cite{Will}, eq. (9.5))
$${\vec \Omega}= \Bigl(\gamma +\frac{1}{2}\Bigr)\sum_{a}({\vec v}-{\vec v}_a)\times
\nabla \frac{Gm_a}{r_ac^2} 
-\frac{1}{2}(\gamma +1)\sum_aG[{\vec
J}_a-3\hat{{\vec n}}_a(\hat{{\vec n}}_a\cdot {\vec J}_a)]/r^3_ac^2-$$
\begin{equation}
-\frac{1}{2}\sum_{a}\vec{v}_a\times \nabla \frac{Gm_a}{r_ac^2},
\label{1.1}
\end{equation}
where $\gamma$ is a PPN parameter which takes value 1 according to the General Relativity, 
$\vec{v}$ is the velocity of the gyroscope, ${\vec v}_a$ is the
velocity of the $a$-th spherical body, ${\vec J}_a$ is its angular
momentum and $r_a$ is its distance to the gyroscope. The third term is anomalous 
since it depends on the velocity of each body \cite{Will}. 
Although there has been an effort to explain why experimentally this term cannot be observed, or that 
it leads to a small periodic effect in case of observation as the Gravity Probe B experiment \cite{Will}, 
it remains to be an anomaly. In the paper \cite{CEJP} this anomaly is solved by assuming axiomatically 
a precession of the coordinate system, i.e. coordinate axes. This precession is analogous to the 
Thomas precession, which is related to 
precession of the gyroscope, and so we will call it Thomas precession too. The final conclusion is that  
observed close from the gyroscope the precession of the gyroscope's axis is given by 
\begin{equation}
{\vec \Omega}_{gyr.}=2\sum_{a}({\vec v}-{\vec v}_a)\times \nabla 
\frac{Gm_a}{r_ac^2} -  
\sum_aG[{\vec J}_a-3\hat{{\vec n}}_a(\hat{{\vec
n}}_a\cdot {\vec J}_a)]/r^3_ac^2, \label{1.2}
\end{equation}
the observed apparent (not true) precession of the distant stars, which is a consequence of the 
precession of the coordinate system, is given by   
\begin{equation}
{\vec \Omega}_{stars}=\frac{1}{2}\sum_{a}({\vec v}-{\vec v}_a)\times
\nabla \frac{Gm_a}{r_ac^2} 
-\frac{1}{4} \sum_aG[{\vec J}_a-3\hat{{\vec
n}}_a(\hat{{\vec n}}_a\cdot {\vec J}_a)]/r^3_ac^2, \label{1.3}
\end{equation}
and hence their subtraction, i.e. the relative precession of the gyroscope's axis
with respect to the distant stars, is given by  
\begin{equation}
{\vec \Omega}_{rel.}=
\frac{3}{2}\sum_{a}({\vec v}-{\vec v}_a)\times
\nabla \frac{Gm_a}{r_ac^2} 
-\frac{3}{4} \sum_aG[{\vec J}_a-3\hat{{\vec
n}}_a(\hat{{\vec n}}_a\cdot {\vec J}_a)]/r^3_ac^2. \label{1.4}
\end{equation}
All these precessions are Lorentz covariant angular velocities. The rotation of the coordinate 
axes is necessary in order to obtain a Lorentz covariant result. But the precession 
(\ref{1.3}) is also necessary if we want to obtain Lorentz covariant results 
about the equations of motion, for some special choices of the observer.      

Notice that according to (\ref{1.4}) the geodetic precession with respect 
to the distant stars is the same as it is well known 
from the General Relativity and as experimentally confirmed by the Gravity Probe B experiment (\cite{GPB-PRL})  
and precession of the system Earth-Moon as a gyroscope around the Sun. But the frame dragging is 25\% less than the 
known values from the General Relativity. Notice that these 25\% cannot be detected via the Lense-Thirring 
effect which arises from the equations of motion, while the frame dragging effect arises from the Fermi-Walker 
connection. So the Gravity Probe B is a unique experiment where these 25\% would be measurable. Unfortunately, 
the large uncertainties in this experiment do not permit precise value for the frame dragging \cite{GPB-PRL}. 
But, the following comments can be of interest. Among the four gyroscopes in the Gravity Probe B experiment let us 
consider the gyroscope no. 3, for which the geodetic precession is the most close to the predicted value 
-6,606.1 mas/yr, which is without doubt the true value. It is natural to expect that the measurements of the 
frame dragging with this gyroscope are also the most close to the true value. 
The measurements with this gyroscope of the frame dragging 
$(-25.0\pm  12.1)$mas/yr are much closer to the value -29.4 mas/yr predicted by (\ref{1.4}), than the 
value -39.2 mas/yr predicted by the General Relativity.     
    
The mentioned anomaly appears as a consequence of the fact that the space-time is a priori 
assumed to be 3+1-dimensional. In \cite{CEJP} the problem is solved 
and equations (\ref{1.2}), (\ref{1.3}) and (\ref{1.4}) are deduced: 

- from the following axiom ``An observer who rests with respect to a non-rotating gravitational body
observes no precession of the coordinate axes of any freely moving coordinate system"
and 

- accepting the known formula for the geodetic precession 
$\frac{3}{2}\sum_{a}({\vec v}-{\vec v}_a)\times \nabla \frac{Gm_a}{r_ac^2}$ from (\ref{1.4}) 
with respect to the distant stars, which is experimentally confirmed.  

\noindent But on the other side, 
since all calculations in differential geometry, General Relativity and tensor calculus are done with respect 
to the chosen coordinate system, but not with respect to the distant stars,  
it is natural to expect that the geodetic precession 
$2\sum_{a}({\vec v}-{\vec v}_a)\times \nabla \frac{Gm_a}{r_ac^2}$ from (\ref{1.2}) should be derived in local 
coordinate system. But this is not a case in the frame of the General Relativity. This is the second anomaly. 

In the papers \cite{IJTP,Tensor1993} is presented a theory of gravitation in flat Minkowski space, 
which fits with all of the gravitational experiments which are performed for verification of the 
General Relativity (deflection of the light rays near the Sun, advance of the perihelion, 
geodetic precession, gravitational radiation and so on).   
Although the first anomaly was solved inside the 3+1-dimensional space-time, the second anomaly cannot be 
solved at the same time. It is solved (in an incoming paper) by researching gravitation in 
3+3-dimensional space-time. The main feature there is that it is not considered parallel transport of vectors, but 
parallel transport of Lorentz transformations. The equations of motion 
in 3+1-dimensional space-time from \cite{IJTP,Tensor1993} can be converted into the 3+3-dimensional space-time. 
Hence the previous predictions remain unchanged and moreover the  
anomalies with the precession of the gyroscope's axis are solved there. 

After all these arguments we see that the space-time is not so simple, and it requires answer of some 
basic questions about the space, time, velocities and motions. Such questions will be subject of this paper. The results 
about the motion of the gyroscopes and change of their spinning axes  
are independent and complementary to the present theory about mechanics of gyroscopes, which is  
based on the Euler's equations and the Lagrange's equation.   

We assume in this paper to use the temporal coordinate as $ict$, instead of $ct$.

\setcounter{equation}{0} 

\section{3-dimensional time}
\label{sec:2}

In this section we give some preliminaries about the concept of the 3-dimensional time 
(\cite{Tensor2009,VB,MB}). 

\subsection{Some basic steps}

We mentioned in the previous section that it is better to consider parallel transport   
of a Lorentz transformation instead of parallel transport of a 4-vector of velocity. 
The result is not the same, because the 
parallel transport of a 4-vector of velocity is always a 4-vector, whereas the parallel transport of 
a Lorentz transformation which is a Lorentz boost at the initial instant, i.e. without spatial rotation, 
may not be a Lorentz boost, because it may contain a spatial rotation too. 
In the case of parallel transport of a 4-vector of velocity,  
the consequence is the appearance of 
anomalies, which were described in the previous section. 
This is the main motivation to research a model of 3-dimensional time. 
Albert Einstein and Henri Poincare many years ago  
thought about 3-dimensional time, such that the space and time would be of the same 
dimension. At present time most of the authors \cite{11,10,1,2,3,4,6,7,8,9} 
propose multidimensional time in order to give a better explanation of the quantum mechanics and the spin. 

Let us denote by $x$, $y$ and $z$ the coordinates in our 3-dimensional space ${\mathbb R}^3$,  
and let us consider the principal bundle over the base ${\mathbb R}^3$, whose fiber is 
$SO(3,{\mathbb C})$ and the (complex) Lie group is $SO(3,{\mathbb C})$ too. 
This bundle will be called {\it space-time bundle}. 
Having in mind that the unit component $O_+^{\uparrow}(1,3)$ of the Lorentz group 
is isomorphic to $SO(3,{\mathbb C})$, this bundle can be considered simply as the 
bundle of all moving orthonormal Lorentz frames.  
The space-time bundle locally can be 
parameterized by the following 9 local coordinates 
$x,y,z$; $x_s,y_s,z_s$; $x_t,y_t,z_t$,  
such that the first 6 coordinates 
parameterize locally the subbundle with the fiber $SO(3,{\mathbb R})$. The local coordinates 
$x_s,y_s,z_s$ are called {\em spatial coordinates}, while $x_t,y_t,z_t$ are called {\em temporal coordinates}.  
So this approach in the Special Relativity is called 3+3+3-dimensional model.   
Indeed, to each body are related 3 coordinates for the position, 3 coordinates for the spatial rotation 
and 3 coordinates for its velocity.   

We will consider the analog of the Lorentz boosts from the 3+1-dimensional space-time.   
The next few assumptions are in accordance with the structure of the group $SO(3,\mathbb{C})$.  
The coordinates $x_s,y_s,z_s,x_t,y_t,z_t$ are functions of $x$, $y$, and $z$, 
and assume that the matrix 
\begin{equation}
V=\left [\matrix{
\frac{\partial x_s}{\partial x} &  
\frac{\partial x_s}{\partial y} & \frac{\partial x_s}{\partial z}\cr  
 & & \cr
\frac{\partial y_s}{\partial x} & \frac{\partial y_s}{\partial y} &
\frac{\partial y_s}{\partial z} \cr  
 & & \cr
\frac{\partial z_s}{\partial x} & \frac{\partial z_s}{\partial y} &
\frac{\partial z_s}{\partial z}\cr }\right ] \label{2.1}
\end{equation}
is symmetric, i.e. $V^T=V$, while the matrix 
\begin{equation}
V^*=\left [\matrix{
\frac{\partial x_t}{\partial x} &  
\frac{\partial x_t}{\partial y} & \frac{\partial x_t}{\partial z}\cr  
 & & \cr
\frac{\partial y_t}{\partial x} & \frac{\partial y_t}{\partial y} &
\frac{\partial y_t}{\partial z} \cr  
 & & \cr
\frac{\partial z_t}{\partial x} & \frac{\partial z_t}{\partial y} &
\frac{\partial z_t}{\partial z}\cr }\right ] \label{2.2}
\end{equation}
is antisymmetric, i.e. $V^{*T}=-V^*$. 

Further let us denote 
$$X=x_s+ix_t, \qquad Y=y_s+iy_t, \qquad Z=z_s+iz_t,$$
such that the matrix 
\begin{equation}
{\cal{V}}=\left [\matrix{
\frac{\partial X}{\partial x} &  
\frac{\partial X}{\partial y} & \frac{\partial X}{\partial z}\cr  
 & & \cr
\frac{\partial Y}{\partial x} & \frac{\partial Y}{\partial y} &
\frac{\partial Y}{\partial z} \cr  
 & & \cr
\frac{\partial Z}{\partial x} & \frac{\partial Z}{\partial y} &
\frac{\partial Z}{\partial z}\cr }\right ] \label{2.3} 
\end{equation}
is Hermitian and ${\cal {V}}=V+iV^*$. 

The antisymmetric matrix $V^*$ depends on 3 variables and its general form can be written as 
\begin{equation}
V^*=\frac{-1}{c\sqrt{1-\frac{v^2}{c^2}}}\left [\matrix{
0 & v_z & -v_y\cr
-v_z & 0 & v_x\cr 
v_y & -v_x & 0\cr }\right ] .\label{2.4}
\end{equation}
According to (\ref{2.4}) we can join to $V^*$ a 3-vector $\vec{v}=(v_x,v_y,v_z)$, which   
transforms as a 3-vector.

It is natural to assume that ${\cal {V}}$ should be presented in the form 
\begin{equation}
{\cal {V}}=e^{iA} =\cos A +i\sin A.\label{2.5}
\end{equation}
Assume that $A$ is an antisymmetric real matrix, which is given by 
\begin{equation}
A=\left [\matrix{0 & -k\cos \gamma & k\cos \beta \cr
k\cos \gamma & 0 & -k\cos \alpha \cr -k\cos \beta & k\cos \alpha & 0\cr }\right ],\label{2.6}
\end{equation}
where $\vec{v}=c(\cos \alpha ,\cos \beta ,\cos \gamma )\tanh (k)$ and 
$(\cos \alpha ,\cos \beta $, $\cos \gamma )$ is a unit vector of the velocity vector.    
As a consequence we obtain 
\begin{equation}
\sin A=\frac{-1}{c\sqrt{1-\frac{v^2}{c^2}}}\left [\matrix{
0 & v_z & -v_y\cr
-v_z & 0 & v_x\cr 
v_y & -v_x & 0\cr }\right ] ,\label{2.7}
\end{equation}
i.e. that $V^*=\sin A$ is given by (\ref{2.4}), 
while the symmetric $3\times 3$ matrix $\cos A$ is given by 
\begin{equation}
(\cos A)_{ij}=V_4\delta_{ij}+\frac{1}{1+V_4}V_iV_j,\label{2.8}
\end{equation}
where 
\begin{equation}
(V_1,V_2,V_3,V_4)=\frac{1}{ic\sqrt{1-\frac{v^2}{c^2}}}(v_x,v_y,v_z,ic). 
\label{2.9}
\end{equation}

According to (\ref{2.2}) and (\ref{2.4}) the temporal vector in this special case is given by 
\begin{equation}
(x_t,y_t,z_t)=\frac{\vec{v}}{c\sqrt{1-\frac{v^2}{c^2}}}\times (x,y,z)+
(x_t^0,y_t^0,z_t^0),\label{2.10}
\end{equation}
where $(x_t^0,y_t^0,z_t^0)$ does not depend on the basic coordinates. 
The coordinates $x_t,y_t,z_t$ are independent and they cover the Euclidean space   
${\mathbb R}^3$ or an open subset of it. But the Jacobi matrix 
$\Bigl [\frac{\partial (x_t,y_t,z_t)}{\partial (x,y,z)}\Bigr ]$ is a singular
as antisymmetric matrix of order 3, where the 3-vector of velocity maps into zero vector. 
So the quantity $(x_t,y_t,z_t)\cdot \vec{v}$ 
does not depend on the basic coordinates and hence we assume that it determines the 1-dimensional time $t$
measured from the basic coordinates.   
Further, one can easily verify that  
$(1-\frac{v^2}{c^2})^{-1/2}(\vec{v}\times (x,y,z))=(1-\frac{v^2}{c^2})^{-1/2}(\vec{v}\times (\cos A)^{-1}(x_s,y_s,z_s)^T)=
\vec{v}\times (x_s,y_s,z_s)$ for simultaneous points in basic coordinates. 
So the formula (\ref{2.10}) becomes  
\begin{equation}
(x_t,y_t,z_t)=\frac{\vec{v}}{c}\times (x_s,y_s,z_s)+
\vec{c}\cdot \Delta t,\label{2.11}
\end{equation}
where $\vec{c}$ is the velocity of light, which has the same direction as $\vec{v}$, i.e. 
$\vec{c}=\frac{\vec{v}}{v}\cdot c$.

\subsection{Local Isomorphism Between $O_{+}^{\uparrow}(1,3)$ and $SO(3,\mathbb{C})$}

Now let us consider the following mapping $F:O_+^{\uparrow}(1,3)\rightarrow SO(3,{\mathbb C})$ given by  
\begin{equation}
\left [\begin{array}{ccc}
M& 0\\ 0& 1\end{array}\right ]\cdot  
\left [\begin{array}{ccccccc}
1-\frac{V_{1}^{2}}{1+V_4} & -\frac{V_{1}V_{2}}{1+V_4} & -\frac{V_{1}V_{3}}{1+V_4}
 & V_{1}\\ 
 & & & \\ 
-\frac{V_{1}V_{2}}{1+V_4} &
1-\frac{V_{2}^{2}}{1+V_4} & -\frac{V_{3}V_{2}}{1+V_4} & V_{2}\\
 & & &\\ 
-\frac{V_{1}V_{3}}{1+V_4} & -\frac{V_{3}V_{2}}{1+V_4} & 1-\frac{V_{3}^{2}}
{1+V_4} & V_{3}\\ 
 & & & \\ 
-V_{1} & -V_{2} & -V_{3} & V_{4}\end{array}\right ] 
\mapsto M\cdot (\cos A+i\sin A),\label{20.3}
\end{equation}
where $\cos A$ and $\sin A$ are given by (\ref{2.8}) and (\ref{2.7}).  
This is well defined because the 
decomposition of any matrix from $O_+^{\uparrow}(1,3)$ as product of a spatial rotation $M$ and a boost 
is unique. Moreover, (\ref{20.3}) is an isomorphism between the two groups and the corresponding isomorphism 
between their Lie algebras is given by 
$$\left [\begin{array}{cccc}
0 & c & -b & ix\\ 
-c & 0 & a & iy\\
b & -a & 0 & iz\\
-ix & -iy & -iz & 0\end{array}\right ]\quad \mapsto 
\left [\begin{array}{ccc}
0 & c+iz & -b-iy\\ 
-c-iz & 0 & a+ix \\
b+iy & -a-ix & 0 \end{array}\right ]. $$
The isomorphism between the groups $O_+^{\uparrow}(1,3)$ and $SO(3,{\mathbb C})$ is the main reason 
why we observe our space-time as 4-dimensional where the group $O_+^{\uparrow}(1,3)$ acts on it, instead of 
3+3-dimensional space-time.     

\subsection{Observation of lengths of a moving body}
We study the observations of lengths in arbitrary direction, 
assuming that there is no effective motion, but simply rotation for an imaginary angle. 
Such a transformation will be called {\em passive motion}. 

The base manifold $\mathbb{R}^3$ is 3-dimensional and it is convenient to consider it as a section of 
$\mathbb{C}^3$, consisting of $(x,y,z,ct_x,ct_y,ct_z)$, where $ct_x=ct_y=ct_z=0$ 
at a chosen initial instant, and we may call it a complex base. The change of the coordinates  
can be done via the $6\times 6$ real matrix 
\begin{equation}
\left [\matrix{ M\cos A& -M\sin A\cr M\sin A&M\cos A\cr }\right ].\label{4.1}
\end{equation}
It acts on the 6-dimensional (tangent) vectors 
$(\Delta x,\Delta y,\Delta z,0,0,0)^T$ of the introduced complex base. 
Multiplying the (tangent) vectors $(\Delta x,\Delta y,\Delta z,0,0,0)^T$
of the complex base from the left with the matrix (\ref{4.1}), 
we obtain 3-dimensional base subspaces as they are 
viewed from the observer who rests with respect to the chosen complex base.   

The quantity $(\Delta x)^2+(\Delta y)^2+(\Delta z)^2$ is invariant, 
where $\Delta x$, $\Delta y$ and $\Delta z$ should be considered as complex quantities 
previously described, because the structure group is $SO(3,\mathbb {C})$. 
Hence we have now two invariants: 
$Re[(\Delta x)^2+(\Delta y)^2+(\Delta z)^2]$ and 
$Im[(\Delta x)^2+(\Delta y)^2+(\Delta z)^2]$.   
According to (\ref{2.11}) the second invariant 
$Im[(\Delta x)^2+(\Delta y)^2+(\Delta z)^2]=2(x_s,y_s,z_s)\cdot (x_t,y_t,z_t)$ is zero for 
simultaneous initial point and end point, i.e. $\Delta t=0$ in (\ref{2.11}). 
But, in case of non-simultaneity ($\Delta t \neq 0)$ the 
situation is different. $Re[(\Delta x)^2+(\Delta y)^2+(\Delta z)^2]$ remains an invariant, but 
$Im[(\Delta x)^2+(\Delta y)^2+(\Delta z)^2]$ is not invariant anymore because of the following reason. In (\ref{2.11}) 
we deduced that the component $\vec{c}\Delta t$ is collinear with the velocity vector $\vec{v}$ because the 
observation is with respect to the basic coordinates $x,y,z$. If the observation is with respect to the coordinates $x_s,y_s,z_s$, 
then $\vec{c}\Delta t$ would be collinear with $(\Delta x_s,\Delta y_s,\Delta z_s)$. So, although $\vec{c}\Delta t$ does not have an 
invariant direction and consequently $Im[(\Delta x)^2+(\Delta y)^2+(\Delta z)^2]$ is not an invariant in this case, we notice that 
$\vec{c}\Delta t$ is always orthogonal to $\frac{\vec{v}}{c}\times (\Delta x_s,\Delta y_s,\Delta z_s)$, 
and hence $\vert \vec{r}_t\vert$ is well defined and $Re[(\Delta x)^2+(\Delta y)^2+(\Delta z)^2]$ is an invariant. 
Note that in case of simultaneity ($\Delta t=0$ in (\ref{2.11})) 
we shall see in section 6 that can be introduced 4 invariants.  
   
Since $Re[(\Delta x)^2+(\Delta y)^2+(\Delta z)^2]$ is invariant, according to the previous notations, it leads to  
\begin{equation}
(\Delta \vec{r}_s)^2-(\Delta \vec{r}_t)^2=(\Delta \vec{r})^2-c^2(\Delta \vec{t})^2,\label{4.2}
\end{equation}
where $\vec{r}_s=(x_s,y_s,z_s)$, $\vec{r}_t=(x_t,y_t,z_t)$, $\vec{r}=(x,y,z)$ and 
$\vec{t}=(t_x,t_y,t_z)$. Hence 
assuming that the initial point and end point are simultaneous with respect to the basic coordinates we should replace  
$\Delta t=0$ in (\ref{2.11}). In basic coordinates it is also $\Delta \vec{t}=(0,0,0)$ and hence we obtain 
$$
(\Delta \vec{r}_s)^2 
-\Bigl (\frac{\vec{v}}{c\sqrt{1-\frac{v^2}{c^2}}}\times \Delta \vec{r}\Bigr )^2=
(\Delta \vec{r})^2,
$$
\begin{equation}
\vert \Delta \vec{r}_s \vert = \vert \Delta \vec{r}\vert 
\sqrt{1+\frac{\frac{v^2}{c^2}\sin^2 \theta}{1-\frac{v^2}{c^2}}},\label{4.3} 
\end{equation}
where $\theta$ is the angle between $\vec{v}$ and $\Delta \vec{r}$. 
Thus in the direction of motion
$(\theta =0)$ there is no observed change in length, while for lengths orthogonal to the motion
$(\theta =\pi /2)$, they are observed larger by $\Bigl (1-\frac{v^2}{c^2}\Bigr )^{-1/2}$ times.   

If there is an {\em active motion}, i.e. {\em there is change of the basic coordinates}, 
according to the previous discussion it is convenient if  
all of the previously described observed lengths in any direction additionally 
should be multiplied by the coefficient $\sqrt{1-\frac{v^2}{c^2}}$. As a consequence the observations for lengths 
for passive and active motions together is in agreement with the classically known results.  

\subsection{Lorentz transformations as transformations on ${\mathbb C}^3$}

For the sake of simplicity we will omit the symbol ``$\Delta$" to denote the difference of coordinates. 
So we assume that the initial point 
of the considered space-time vector has coordinates equal to zero. 
Assume that $x,y,z$ are basic coordinates. Let the coordinates $x_s,y_s,z_s$ be  
denoted by $x',y',z'$ and let us denote $\vec{r}=(x,y,z)$ and $\vec{r'}=(x',y',z')$. 
It is of interest to see the form of the Lorentz boosts as transformations in ${\mathbb C}^3$, while the spatial rotations  
are identical in both cases.  

{\bf Theorem.} {\em 
The following transformation in ${\mathbb C}^3$ 
\begin{equation}
\Bigl (1-\frac{v^2}{c^2}\Bigr )^{-1/2}
\left [\matrix{\vec{r'}\cr \vec{c}t'+\frac{\vec{v}\times \vec{r'}}{c}\cr }\right ] 
=\left [\matrix{ \cos A& -\sin A\cr \sin A & \cos A\cr }\right ]
\left [\matrix{\vec{r}+\vec{v}(t+\delta t)\cr \vec{c}(t+\delta t)\cr }\right ]\label{7.1}
\end{equation}
via the group $SO(3,{\mathbb C})$ where $\delta t=\frac{\frac{\vec{r'}\vec{v}}{c^2}}{\sqrt{1-\frac{v^2}{c^2}}}$, 
is equivalent to the transformation of a Lorentz boost determined by the isomorphism (\ref{20.3}). } 

The coefficient $\beta =(1-\frac{v^2}{c^2})^{-1/2}$
is caused by the active motion from subsection 2.3. Obviously we have translation in the basic coordinates for vector 
$(\vec{v}(t+\delta t),\vec{c}(t+\delta t))$, where 
$\delta t=\frac{\vec{r'}\vec{v}}{c^2}/\sqrt{1-\frac{v^2}{c^2}}$ 
appears from the non-simultaneity of the start point and the endpoint analogously as in the Special Relativity.   
On the other side, 
according to (\ref{2.11}) in the moving system we have the temporal vector $\vec{v}\times \vec{r'}/c$, 
which disappears in basic coordinates ($\vec{v}=0$). 
  
According to the previous theorem 
the well known 4-dimensional space-time is not fixed in 6 dimensions, but changes with the direction 
of velocity. Namely this 4-dimensional space-time is generated by the basic space vectors and the velocity vector
from the imaginary part of the complex base.    

Having in mind the equation (\ref{2.11}), the Lorentz transformation (\ref{7.1}) can be written 
in the following form 
\begin{equation}
\Bigl (1-\frac{v^2}{c^2}\Bigr )^{-1/2}
\left [\matrix{\vec{r}_s \cr \vec{r}_t\cr }\right ]
=\left [\matrix{ \cos A& -\sin A\cr \sin A & \cos A\cr }\right ]
\left [\matrix{\vec{r}+\vec{v}(t+\delta t)\cr \vec{c}(t+\delta t)\cr }\right ].\label{7.2}
\end{equation}
Since the coordinates $x_s,y_s,z_s,x_t,y_t,z_t$ are measured according to the 
basic coordinates $x,y,z$, we know that all of them are observed contracted for coefficient 
$\sqrt{ 1-\frac{v^2}{c^2}}$ so we need to multiply them with  $\beta =(1-\frac{v^2}{c^2})^{-1/2}$. If we denote 
again the same coordinates but now measured from the self coordinate systems, then the Lorentz transformation becomes 
\begin{equation}
\left [\matrix{\vec{r}_s \cr \vec{r}_t\cr }\right ]=
\left [\matrix{ \cos A& -\sin A\cr \sin A & \cos A\cr }\right ]
\left [\matrix{\vec{r}+\vec{v}(t+\delta t)\cr \vec{c}(t+\delta t)\cr }\right ].\label{7.3}
\end{equation}

\setcounter{equation}{0} 
\section{Introduction to the 3+3+3-model of cosmology}
\label{sec:3}

Now let us introduce a possible global construction of the universe. 
At the beginning of the previous section we considered the space-time bundle with base $B$ and fiber 
$SO(3,{\mathbb C})$. The group $SO(3,{\mathbb C})$ is very convenient because it can be considered 
simultaneously as a fiber and Lie group. This is not the case for its isomorphic group 
$O_+^{\uparrow}(1,3)$, which acts on 4-dimensional space-time. 
For the sake of simplicity we assumed that $B={\mathbb R}^3$, but the basic space 
topologically should be closed and bounded as it is accepted and so it should be changed. On the other hand 
the equality (\ref{7.3}) suggests that the space-time (without rotations) is a complex manifold. 
Indeed, analogously as $(x+ict_x,y+ict_y,z+ict_z)$ is a local coordinate neighborhood of $SO(3,{\mathbb C})$, 
$(x_s+ix_t,y_s+iy_t,z_s+iz_t)$ is also a coordinate neighborhood of the same
manifold. These two coordinate systems can 
be considered as coordinate neighborhoods of $SO(3,{\mathbb C})$ as a complex manifold, because    
according to (\ref{7.3}) the Cauchy-Riemannian conditions for these two systems are satisfied. 
   
If we assume that the basic space is homeomorphic to $RP^3$, i.e. $SO(3,{\mathbb R})$, we arrive at a satisfactory result. 
The complexification of this group is $SO(3,{\mathbb C})$ and it is the same as the fiber and the Lie group of the 
space-time bundle. 
The local coordinates of $SO(3,{\mathbb R})$ are angles, i.e. real numbers, but we use length units for 
our local spatial coordinates. So each small angle $\varphi$ of rotation in a given direction corresponds to a coordinate length 
$R\varphi$ in the same direction, where $R$ is a constant, which can be called the {\em radius of the universe}.  
We have a similar situation if we assume that the basic space is homeomorphic to $S^3$, i.e. the Lie group of unit quaternions. 
 
We give now some additional reasons for this assumption. Both topological 
spaces $SO(3,{\mathbb R})$ and $S^3$ admit 3 linearly independent vector fields, which are orthogonal at 
each point, because each $n$-dimensional real Lie group admits $n$ linearly independent vector 
fields. Indeed we can choose $n$ vector fields at the unit and then we may parallel transport 
using the group structure to any other point. 
Each Lie group admits a connection with zero curvature, but non-zero 
torsion tensor in the general case. The previously mentioned three vector fields may be 
chosen to be orthonormal at each point, and they are parallel with respect to the mentioned 
connection with zero curvature. This is in accordance with the 
recent astronomical observations, which show that our universe is a flat (non-curved) space.  

Let us discuss the space-time dimensionality of the universe. 
We have that it is parameterized by the following 9 independent 
coordinates: $x,y,z$ coordinates which locally parameterize the spatial part of the universe $SO(3,{\mathbb R})$ or $S^3$, 
and $x_s,y_s,z_s,x_t,y_t,z_t$ coordinates which parameterize the fiber.  
Also,     
the partial derivatives of these coordinates with respect to $x,y,z$ lead to the same manifold, but now as 
a group of transformations. So in any case the total space-time of the universe 
is homeomorphic to $SO(3,{\mathbb R})\times SO(3,{\mathbb C})$, i.e. $SO(3,{\mathbb R})\times {\mathbb R}^3\times SO(3,{\mathbb R})$, or
$S^3\times {\mathbb R}^3\times S^3$. 
In the above parameterization, ${\mathbb R}^3$ is indeed the space of velocities such that $\vert \vec{v}\vert <c$. The group 
$SO(3,{\mathbb C})$, as well as its isomorphic group $O_+^{\uparrow}(1,3)$, considers only velocities with magnitude less than $c$. 
If $\vert \vec{v}\vert =c$, then we have a singularity. 

Notice that if we consider that the universe is a set of points, then it is natural 
to consider it as 6-dimensional. But in such 6-dimensional space-time, rotations will not be admitted, because the 
existence of the 3-dimensional space does not mean that the spatial rotations are also admitted. Analogously, if we neglect the three temporal coordinates, according to (\ref{2.11}) the velocities will not be admitted. 
But, since we consider the universe as a set of orthonormal moving frames, so it is 
more natural to consider it as 9-dimensional.      

This 9-dimensional space-time has the following property: From each point of the space, each velocity and each spatial direction of 
the observer, the universe seems to be the same. In other words, assuming that $R$ is a global constant,
there are no privileged spatial points,   
no privileged directions and no privileged velocities.   

We can resume that   
this 3+3+3-model consists of three 3-dimensional sets: velocity (V) which is homeomorphic to 
${\mathbb R}^3$, space (S) which is homeomorphic to $SO(3,{\mathbb R})$ or $S^3$, 
and spatial rotations (SR) which is homeomorphic to 
$SO(3,{\mathbb R})$ or $S^3$. We assume that the spaces $S$ and $SR$ are homeomorphic because of symmetry. So, if we assume that $S=S^3$, then $SR$ should be also $S^3$ as a universal covering group of $SO(3,{\mathbb R})$. 

For the sake of simplicity we shall use mainly the group $SO(3,{\mathbb R})$ instead of $S^3$. Indeed, they are locally isomorphic.  

\setcounter{equation}{0} 
\section{Research of the 6-dimensional space $SR\times S$}
\label{sec:4}

In this section and also in the following sections 
we will investigate the Lie groups in the 3+3+3-model and their invariants. 
We shall see that there is a symmetry between the sets $S$ and $SR$. 

We almost know from the Special Relativity that $SR\times V\cong SO(3,{\mathbb R})\times {\mathbb R}^3$ is indeed the Lorentz group 
$O_{+}^{\uparrow}(1,3)$ which 
is isomorphic to $SO(3,{\mathbb C})$. 
According to (\ref{7.3}) the differential of any mapping from $SO(3,{\mathbb C})$ can be treated as a Lie group and fiber over any point of $S$, 
because $\left [\matrix{\vec{r}_s \cr \vec{r}_t\cr }\right ]$   
and $\left [\matrix{\vec{r}+\vec{v}(t+\delta t)\cr \vec{c}(t+\delta t)\cr }\right ]$   
correspond to the same point from the base in two coordinate neighborhoods.   
So we consider this group for a fixed basic point in $S$. 
It means that the product $SR\times V\cong SO(3,{\mathbb C})$ can 
be considered as a fiber and Lie group of a trivial principal bundle with base $S$. 
The Lie group $SO(3,{\mathbb C})$ will be denoted by $G_t$, with index $t$ (time). 
This principal bundle is close to the methods of classical mechanics. Indeed, it is sufficient 
to study the law of the change of the matrices from the fiber, i.e. the matrices which contain the information about the spatial rotation and the 
velocity vector of a considered test body. Then using integration it is easy to find the trajectory of motion of the test body. 

Now let us consider the product $S\times SR$.  
The product $S\times SR$ can 
be considered as a fiber and Lie group of a trivial principal bundle over the base $V$. So we consider this group for a fixed temporal point in $V$, 
i.e. 
for a fixed inertial coordinate system up to a translation and spatial rotation. 
So the coefficient $\sqrt{1-\frac{v^2}{c^2}}$ will not have any role. It doesn't mean that 1-parametric 
time does not change, but simply the velocities may be very slow and neglected.    
The structure Lie group will be denoted by $G_s$ with index $s$ (space), and this group should be analogous to the 
group of all rotations and translations in the 3-dimensional Euclidean space. On the other side this group is analogous to the 
Lorentz group $G_t$. While the Lie algebra of $G_t$ has the form 
\begin{equation}
\left [\begin{array}{ccc}C & &B \\ \\ -B& & C\end{array}\right ],
\end{equation} 
the Lie algebra of $G_s$ is given by 
\begin{equation}
\left [\begin{array}{ccc}C & &B \\ \\ B& & C\end{array}\right ].\label{12.1}
\end{equation}   
Here $C$ and $B$ are antisymmetric $3\times 3$ matrices, where the matrix $C$ is the Lie algebra 
which corresponds to the spatial rotations, i.e. to the Lie algebra of $SO(3,{\mathbb R})$ and $S^3$, 
while $B$ from (\ref{12.1}) is an antisymmetric matrix 
which corresponds to the Lie algebra of $S$, if we identify $S$ with the Lie group $SO(3,{\mathbb R})$ or $S^3$. So, if we put  
$$
C=\left [\begin{array}{ccc}0 & -c_3 & c_2\\ c_3 & 0 & -c_1 \\ -c_2 & c_1 & 0\end{array}\right ], \qquad 
B=\left [\begin{array}{ccc}0 & -b_3 & b_2\\ b_3 & 0 & -b_1 \\ -b_2 & b_1 & 0\end{array}\right ],$$
$(c_1,c_2,c_3)$ determines a 3-vector for a small spatial rotation, 
while $(b_1,b_2,b_3)$ is proportional to the 3-vector of spatial translation   
in $S$. In a special case the coefficient of proportionality may be $1/R$, where $R$ is the radius of the universe.
Since $1/R$ is extremely small quantity, $R(b_1,b_2,b_3)$ can be considered as a vector of translation.    
Other cases with much larger coefficient of proportionality will be considered later. 

It is easy to verify that the mapping 
$$\left [\begin{array}{ccc}C & &B \\ \\ B& & C\end{array}\right ]\quad \mapsto \quad (C+B,C-B),$$
defines an isomorphism between the Lie algebra of $G_s$ and the Lie algebra of $SO(3,{\mathbb R})\times SO(3,{\mathbb R})$,
i.e. $S^3\times S^3$. 
Note that $B \mapsto -B$ gives an automorphism of the group $G_s$.   
These comments support our assumption from section 3 that $S$ is homeomorphic to $SO(3,{\mathbb R})$ or $S^3$, 
such that $S\times SR$ is really homeomorphic to 
$SO(3,{\mathbb R})\times SO(3,{\mathbb R})$ or $S^3 \times S^3$. 
  
While the matrices from the group $G_t=SO(3,{\mathbb C})$ can be interpreted in the 3+1-dimensional space-time, the matrices 
from the group $G_s$ do not have such interpretation in 3+1-dimensional space-time.  
But if we neglect the terms of order $1/R^2$, the group $G_s$ reduces to the group 
of all rotations and translations in the Euclidean space, i.e. all matrices of type 
$\left [\begin{array}{ccc}M & &\vec{h}^T\\ \\0& & 1\end{array}\right ]$, 
where $M\in SO(3,{\mathbb R})$ 
and ${\vec{h}}^T$ is the vector of translation. Indeed, the corresponding mapping for 
their Lie algebras appears to be approximately an isomorphism (neglecting $1/R^2$) between the corresponding Lie algebras. 
Analogously if we neglect $c^{-2}$, then the group $G_t$ reduces to the group of Galilean transformations. 

We will determine the set of matrices in $G_s$ in the following way. We know that any matrix from $G_t$ can be written as a product 
of a Lorentz boost and a spatial rotation, i.e. matrix of form 
$\left [\begin{array}{ccc}M & &0\\ \\0& & M\end{array}\right ]$, 
where $M$ determines a 
spatial rotation, and conversely, as a product of a spatial rotation and a Lorentz boost. Analogously it can also be proved for the 
group $G_s$. So we need only to determine the matrices which are analogous to the Lorentz boosts. This subgroup of transformations 
which correspond to the "translations" in the $z$-direction are given by matrices of the form    
\begin{equation}
\left [\begin{array}{cccccc}\cos \alpha & 0 & 0 & 0 & \sin\alpha & 0\\
0 & \cos \alpha & 0 & -\sin\alpha & 0 & 0\\
0 & 0 & 1 & 0 & 0 &0 \\
0 & \sin \alpha & 0 & \cos\alpha & 0 & 0\\
-\sin \alpha & 0 & 0 & 0 & \cos\alpha & 0\\
0 & 0 & 0 & 0 & 0 & 1 \end{array}\right ]. \label{12.2}
\end{equation}

Although $G_s$ is isomorphic  
to the Cartesian product $SO(3,{\mathbb R})\times SO(3,{\mathbb R})$ or $S^3\times S^3$, the two multipliers in this isomorphism are 
not $S$ and $SR$. 

This principal bundle will have important applications in section 5. 
If a {\em rigid body} is spinning, there may appear a constraint for 
the spatial rotation, but no constraint for the place position of the moving frame. Before we consider it in more details, we give 
examples when these constraints occur.   

{\bf Example 1.}  Assume that we have a rigid circle which rotates around its axis. Then it appears a constraint for the particles of the 
circle, because the circle is a rigid body and the spatial rotation for all particles must be the same. 
So the Thomas precession which appears for particles of the circle cannot be realized.   

{\bf Example 2.} While the previous example leads to a negligible relativistic effect, this example gives easy observable effects. 
We know that spinning bodies, like coins and so on, 
where the gyroscope's axis is not constant, show some departures from the classical physical laws, for example the angular momentum is not 
preserved. 

In case of a constraint, the 3-vector of unadmitted 
angular rotation will be converted into spatial displacement by vector multiplication with the corresponding vector $\vec{r}$, 
whose center is the center of the osculating circle.     
This is general conclusion and later in section 6 it will be supported by the 6-dimensional space $S\times SR$ with the structural group $G_s$. 
This property of conversion from unadmitted spatial rotation into spatial displacement will be called {\em basic property of the space}.  
This displacement in the space which is a result of the basic property of the space will be called {\em spin displacement}. Its first and 
second derivative by the 1-parametric time $t$ will be called {\em spin velocity} and {\em spin acceleration} and will be 
denoted by large letters $V$ and $A$ respectively. 
While the ordinary, i.e. inertial velocity has the property of inertia, the spin velocity is non-inertial,  
because it can be conceived just like a displacement in space.

\setcounter{equation}{0} 
\section{Application to the rotating bodies} 
\label{sec:5}

In this section we assume that there is not any gravitational acceleration or any electromagnetic interaction. 
First we shall study a rotating sphere with radius $R$. 
This is uniquely determined by a family of orthogonal matrices $A(t)\in SO(3,{\mathbb R})$, 
where $t$ is a 1-dimensional temporal parameter.  
We assume that $A(t)$ is a set of differentiable matrices of sufficient order and $A'(t)\neq 0$. 
At each instant each point of the sphere moves with a tangent 3-vector of velocity. 
This vector field of velocities is continuous, and so there exists at least one point $X$ on the sphere whose velocity is zero. Then its 
antipode point $X^*$ also has zero velocity vector. Moreover, if there is a third point with zero velocity vector, 
then each point of the sphere has a zero velocity vector, which is in a contradiction to our assumption $A'(t)\neq 0$. 
So at each instant the 
rotating axis $XX^*$ is unique and well defined. Moreover, this axis changes continuously and also the vector of angular velocity 
$\vec{w}$ is well defined at each instant and it is a continuous function. 

Let us choose an arbitrary point of the sphere and let us denote by $\vec {t}$, $\vec{n}$ and $\vec{b}$ the orthonormal 
moving trihedron consisting of the tangent vector, normal vector and binormal vector along its trajectory. 
If there is complete freedom of the spatial rotations 
of each point, then according to the Frenet equations, for the differential of the mapping at the considered point 
with respect to the moving trihedron $(\vec{t},\vec{n},\vec{b})$, we have 
the following antisymmetric matrix 
$$\left [\begin{array}{ccc} 0 & k & 0 \\ -k & 0 & \tau \\ 0 & -\tau & 0\end{array}\right ]ds$$
from the Lie algebra of $SO(3,{\mathbb R})$. Here $k$ is the curvature, $\tau$ is the torsion of the trajectory of the considered point 
and $s$ is the natural parameter.    
On the other side this infinitesimal spatial rotation is not completely permitted, 
because the sphere is a rigid body. We assume that the angular velocity $w$ of the sphere is such that the spinning axis of the sphere 
is parallel or almost parallel to the binormal vector of the trajectory of the considered point. 
So, for this reason we will always assume that $w$ is sufficiently large compared with $\vert \frac{d\vec{b}}{dt}\vert$.  
Then the differential of the matrix $A(t)$ with respect to almost the same 
moving trihedron $(\vec{t},\vec{n},\vec{b})$ is given by 
$$\left [\begin{array}{ccc} 0 & k_0 & 0 \\ -k_0 & 0 & 0 \\ 0 & 0 & 0\end{array}\right ]ds,$$
where $k_0=\frac{wdt}{ds}$. 
This means that the following orthogonal transformation 
$$
\Bigl (\left [\begin{array}{ccc} 1 & 0 & 0 \\ 0 & 1 & 0 \\ 0 & 0 & 1\end{array}\right ]+
\left [\begin{array}{ccc} 0 & k & 0 \\ -k & 0 & \tau \\ 0 & -\tau & 0\end{array}\right ]ds\Bigr )
\Bigl (\left [\begin{array}{ccc} 1 & 0 & 0 \\ 0 & 1 & 0 \\ 0 & 0 & 1\end{array}\right ]+
\left [\begin{array}{ccc} 0 & k_0 & 0 \\ -k_0 & 0 & 0 \\ 0 & 0 & 0\end{array}\right ]ds\Bigr )^{-1}$$
$$\approx \left [\begin{array}{ccc} 1 & 0 & 0 \\ 0 & 1 & 0 \\ 0 & 0 & 1\end{array}\right ]+
\left [\begin{array}{ccc} 0 & k-k_0 & 0 \\ -(k-k_0) & 0 & \tau \\ 0 & -\tau & 0\end{array}\right ]ds
$$
is ``not permitted". Since the matrix 
$$\left [\begin{array}{ccc} 0 & (k-k_0)ds & 0 \\ -(k-k_0)ds & 0 & \tau ds\\ 0 & -\tau ds & 0\end{array}\right ]$$ 
corresponds to the vector $-(k-k_0)\vec{b}ds-\tau\vec{t}ds$,  
according to the basic property of the space, we come to the conclusion that at the considered instant the spin displacement is given by 
\begin{equation}
d\vec{L}=-((k-k_0)\vec{b}ds+\tau \vec{t}ds) \times \vec{r}
=((k-k_0)\vec{b}ds+\tau \vec{t}ds) \times (r\vec{n})=
\tau r\vec{b}ds-(k-k_0)r\vec{t}ds\label{6.1}
\end{equation}
where $r=1/k$ is the radius of the curvature. 
In section 7 we will show that the term $(k-k_0)r\vec{t}ds$ has no role, except in case of a spinning body in a gravitational field.  
So we can write $d\vec{L}/ds$ in the following form 
\begin{equation}
\frac{d\vec{L}}{ds}=\Bigl (\frac{d\vec{b}}{ds}\cdot \vec{n}\Bigr )\vec{t}\times \vec{r}.\label{6.2} 
\end{equation}

{\bf Remark 1.} Note that the formula (\ref{6.2}) is not Lorentz covariant. Indeed, according to 
a moving coordinate system with a constant 3-vector of velocity ${\vec u}$ the curve becomes $\vec{r}-\vec{u}t$ and 
its torsion and curvature are not the same as previous. So the choice of the coordinate system where the formula (\ref{6.2}) holds 
should be precisely determined. We 
a priori assume that the basic property of the space may be applied only in such a system, 
in which the center of the sphere rests. 
Further we can consider arbitrary rotating bodies, not only spheres. Then the condition of the ``system 
where the center of the sphere rests" should be replaced by ``system where the center of the osculating sphere rests". 
In all further formulas this choice of the coordinate system will be assumed without emphasizing it. 
The osculating sphere means the sphere determined by 4 ``infinitesimally close" points near the considered point. 
Let $A$, $A'$, $A''$ and $A'''$ be 4 infinitesimally close points on a trajectory parameterized by the time $t$, and let $A_0$ 
be the center of the sphere which passes through these 4 points. Let after time $\Delta t \approx 0$ we choose another 4 
infinitesimally close points 
$B$, $B'$, $B''$ and $B'''$ and let $B_0$ be the center of the sphere which passes through them. Then in the system 
which moves approximately with velocity $\overrightarrow{A_0B}_0/\Delta t$ the osculating sphere rests and the spin velocity 
should be calculated there. 
In the general case 
the spin velocity should be calculated for each particle of the body in a different system at a chosen instant. 
In this paper all such systems are the same with the system where the barycentre rests. 
The calculated values of $\vec{L}$, $\vec{V}$ or $\vec{A}$ at the 
chosen instant are considered as invariant values from the choice of the system, 
by neglecting the relativistic terms of order $c^{-2}$.      

Now the spin velocity should be integrated over all points of the sphere, in order to get the
averaged spin velocity of the whole sphere. We suppose that the sphere is homogeneous, i.e. 
with the same matter density. 
For the sake of simplicity at the considered instant we assume that  
$\vec{b}=(0,0,1)$, $\vec{t}=(-\sin \theta ,\cos \theta ,0)$, 
$\vec{n}=(-\cos \theta ,-\sin \theta ,0)$, $\vec{r}=(\sqrt{1-z^2}\cos \theta ,\sqrt{1-z^2}\sin \theta ,0)$
and $\frac{d\vec{b}}{dt}=(p,q,0)$. 
Hence we obtain 
$$\int \vec{V}dP = 
-\int [(p,q,0)\cdot (\cos \theta ,\sin \theta ,0)](-\sin \theta ,\cos \theta ,0)\times \vec{r}dP =$$
$$=-\int (p\cos \theta +q\sin \theta )(-\sin \theta ,\cos \theta ,0)\times \vec{r}dP=$$
$$=\int (p\cos \theta +q\sin \theta )(0,0,\sqrt{1-z^2})dP=(0,0,0).$$
The third coordinate is zero, because for each $z_0$ the integration over the circle $x^2+y^2+z^2=R^2,\; z=z_0$ is equal to zero. 

Now let us consider a finite number of points on the sphere with radius vectors $\vec{R}_i$, $i\in\{ 1,\cdots ,n\}$, such that their 
barycentre 
is the center of the sphere, i.e. $\sum_im_i\vec{R}_i=0$. 
First let us prove the following statement: {\em If the distances between arbitrary two points among the considered $n$ points on the  
sphere remain unchanged and if all the trajectories have the same binormal vectors, 
then the spin velocity is identically equal to zero.}   

According to (\ref{6.2}) the spin velocity of the $i$-th body is given by 
$$\vec{V}_i=\frac{d\vec{L}_i}{dt}=\Bigl (\frac{d\vec{b}_i}{dt}\cdot \vec{n}_i\Bigr )\vec{t}_i\times (-r_i\vec{n}_i),$$
\begin{equation}
\vec{V}_i=-\Bigl (\frac{d\vec{b}_i}{dt}\cdot \vec{n}_i\Bigr )r_i\vec{b}_i.\label{6.3}
\end{equation}
Since all the trajectories have the same binormal vectors, it means that for each $i$, $\vec{b}_i=\vec{b}$. 
According to the equality $\sum_i m_i\vec{R}_i=0$, its projection to the spinning axis, which passes through the barycentre, 
leads to the following equality $\sum_i m_ir_i\vec{n}_i=0$, and hence 
$$\vec{V}=\sum_i\vec{V}_i\frac{m_i}{M}=-\sum_i \Bigl (\frac{d\vec{b}}{dt}\cdot \vec{n}_i\Bigr )r_i\vec{b}\frac{m_i}{M}
=-\frac{1}{M}\Bigl (\frac{d\vec{b}}{dt}\cdot\sum_im_ir_i\vec{n}_i\Bigr )\vec{b}=0,$$ 
where $M=\sum_im_i$. 

In the previous statement the points may not lie on one sphere, but it is important that the distances between the points remain unchanged. 
Consequently, {\em if we consider a rigid body spinning around its barycentre 
where all the trajectories have the same binormal vectors, then the spin velocity is identically equal to zero.} 
Notice also that if the trajectories do not have the same binormal vectors, but $w$ is very large, then the spin velocity 
is close to zero, because all the trajectories have almost equal binormal vectors. 

{\bf Example 1.} 
Let us consider a finite number $n$ of bodies with radius vectors $\vec{R}_1,\cdots $, $\vec{R}_n$
and masses $m_1,\cdots ,m_n$ respectively, such that  
$\sum_i \vec{R}_im_i=0$, in order their barycentre to be at the coordinate origin. 
Suppose that these bodies with small dimensions are moving such that 
$\sum_i \vec{R}_im_i\equiv 0$ and the distances between the bodies are not constants in the general case. 
The bodies are connected by rods of constant lengths only to their barycentre which is at the coordinate origin. 
The lengths $R_1=\vert \vec{R}_1\vert ,...,R_n=\vert \vec{R}_n\vert$ are constants, which are not necessarily equal. 
We may imagine that each body is attached to a spinning sphere with mass 0.       
Each imagined sphere may rotate with large angular velocity and the corresponding binormal vector changes slowly. 
So the previously derived formula (\ref{6.3}) 
may be applied to each body separately, because the spin velocity for a trajectory on a spinning sphere 
depends only on the trajectory of the body.  
Finally, we come to the conclusion 
that the acceleration of the whole system of bodies, which are connected to their barycentre, is equal to 
$\vec{A}=\frac{1}{M}\sum_i \frac{d\vec{V}_i}{dt}m_i$, where $M=m_1+\cdots +m_n$ and 
$\vec{V}_i$ is the spin velocity of the $i$-th body, given by (\ref{6.3}). 
In the general case $\vec{A}$ is a non-zero vector, because the binormal vectors of the trajectories are not equal in the general case. 
Hence we have a violation of Newton's third law in its classical formulation.                 

\setcounter{equation}{0} 
\section{Structural invariants for the 3+3+3-model}
\label{sec:6}

In this section we shall verify the basic property of the space via the 6-dimensional space $S\times SR$ and the Lie group $G_s$, which is 
a subgroup of $SO(6,\mathbb{R})$.  
We know that the Lie group $G_s$ has 6 parameters and denote them by 
$\xi_1$, $\xi_2$, $\xi_3$, $\eta_1$, $\eta_2$,$\eta_3$, such that the Lie algebra is parameterized by
$$
\left [\begin{array}{cccccc} 
0 & -\xi_3 & \xi_2 & 0 & \eta_3 & -\eta_2 \\ 
\xi_3 & 0 & -\xi_1 & -\eta_3 & 0 & \eta_1 \\
-\xi_2 & \xi_1 & 0 &  \eta_2 & -\eta_1 & 0 \\
0 & \eta_3 & -\eta_2 & 0 & -\xi_3 & \xi_2\\
-\eta_3 & 0 & \eta_1 & \xi_3 & 0 & -\xi_1\\
\eta_2 & -\eta_1 & 0 & -\xi_2  & \xi_1 & 0\end{array}\right ].
$$
Since the mapping 
$$\left [\begin{array}{ccc}C & &B \\ \\ B& & C\end{array}\right ]\quad \mapsto \quad (C+B,C-B),$$
where $B$ and $C$ are antisymmetric matrices,
defines an isomorphism between the Lie algebra of $G_s$ and the Lie algebra of $SO(3,{\mathbb R})\times SO(3,{\mathbb R})$, this implies that 
we have two invariants: 
$$I_1=[(d\xi_1,d\xi_2,d\xi_3)+(d\eta_1,d\eta_2,d\eta_3)]^2,$$
and 
$$I_2=[(d\xi_1,d\xi_2,d\xi_3)-(d\eta_1,d\eta_2,d\eta_3)]^2.$$
Hence 
\begin{equation}
J_1=\frac{I_1+I_2}{2}=d\xi_1^2+d\xi_2^2+d\xi_3^2+d\eta_1^2+d\eta_2^2+d\eta_3^2\label{70.1}
\end{equation}
and 
\begin{equation}
J_2=\frac{I_1-I_2}{4}=d\xi_1\cdot d\eta_1+d\xi_2\cdot d\eta_2+d\xi_3\cdot d\eta_3\label{70.2}
\end{equation}
are also invariants.   

{\bf Example 1.} 
The vector $(d\eta_1,d\eta_2,d\eta_3)$ is interpreted as spatial displacement, commonly denoted by $(dx,dy,dz)$, 
while the $3\times 3$-matrix  
$$
\left [\begin{array}{ccc} 
0 & d\xi_3 & -d\xi_2 \\ 
-d\xi_3 & 0 & d\xi_1 \\
d\xi_2 & -d\xi_1 & 0 \end{array}\right ]
$$
is indeed the Frenet antisymmetric matrix $\left [\begin{array}{ccc} 0 & k & 0 \\ -k & 0 & \tau \\ 0 & -\tau & 0\end{array}\right ] rds$, 
with respect to the orthonormal frame $(\vec{t},\vec{n},\vec{b})$, 
assuming that the center of the osculating sphere rests (Remark 1 in section 5). 
So with respect to this orthonormal frame, 
$$(d\eta_1,d\eta_2,d\eta_3)=\vec{t}ds \quad \hbox{and}\quad  
(d\xi_1,d\xi_2,d\xi_3)=(r\tau\vec{t}+kr\vec{b})ds=(r\tau\vec{t}+\vec{b})ds.$$ 
Hence for the invariants $J_1$ and $J_2$ we get 
$$J_1=(2+r^2\tau^2)ds^2, \quad J_2=r\tau ds^2.$$
Now, according to the basic property of the space, if the spatial rotation for angle $-\tau \vec{t}ds$ is not permitted, 
as in case of rotating sphere, then we have spatial displacement 
$-\tau\vec{t}ds \times (-r\vec{n})=r\tau \vec{b}ds$. So, the new values of the vectors 
$(d\eta_1,d\eta_2,d\eta_3)$ and $(d\xi_1,d\xi_2,d\xi_3)$ 
become 
$(d\eta_1,d\eta_2,d\eta_3)=\vec{t}ds+r\tau \vec{b}ds$   
and $(d\xi_1,d\xi_2,d\xi_3)=\vec{b}ds$.  
Now it is easy to see that the values of $J_1$ and $J_2$ remain unchanged after this replacement, which confirm the proposed theory. 
In section 8 this Example will be reconsidered without any approximation.  
     
{\bf Example 2.} 
Let us consider a circle which rotates in its plane around its center. If the particles of the circle may freely 
rotate around their axes  according to the Thomas precession, then we have 
$$(d\eta_1,d\eta_2,d\eta_3)=\vec{t}ds, \quad (d\xi_1,d\xi_2,d\xi_3)=\vec{b}ds + \frac{\vec{w}_{Th.}}{w}ds,$$ 
where $\vec{w}_{Th.}=\frac{-1}{2c^2}(\vec{v}\times \vec{a})$ is the Thomas precession. 
On the other hand, if the circle is a rigid body such that the Thomas precession is not permitted to be realized, then according to the 
basic property of the space  
$(d\eta_1,d\eta_2,d\eta_3)=\vec{t}ds+\vec{V}dt$ and   
$(d\xi_1,d\xi_2,d\xi_3)=\vec{b}ds$,  
where $\vec{V}=\vec{w}_{Th.}\times \vec{r}=\frac{-1}{2c^2}(\vec{v}\times \vec{a})\times \vec{r}$. 
Now it is easy to verify that $J_2=0$ in both cases. Having in mind 
that $ds=vdt$, $v=rw$, $a=rw^2$ and the directions of $\vec{w}_{Th.}$ and $\vec{V}$, in both cases we obtain 
$J_1=1+\Bigl [1-\frac{r^2w^2}{2c^2}\Bigr ]^2$. Hence, we have again confirmation that $J_1$ and $J_2$ are really invariants. 

{\bf Example 3.}
Now assume that instead of the Thomas precession we have 
an angular velocity which is opposite to the angular rotation of the circle. Then in the first case, when the particles of the circle may freely 
rotate around their axes, we have 
$$(d\eta_1,d\eta_2,d\eta_3)=\vec{t}ds, \qquad (d\xi_1,d\xi_2,d\xi_3)=\vec{b}ds -\vec{b}ds =0.$$ 
Further if the circle is a rigid body and the rotations of the particles around their axes are not permitted, we have 
$$(d\eta_1,d\eta_2,d\eta_3)=\vec{t}ds+\Bigl (\frac{-\vec{b}ds}{r}\Bigr )\times (-r\vec{n})=0 \quad  
\hbox{and} 
\quad (d\xi_1,d\xi_2,d\xi_3)=\vec{b}ds.$$ 
We notice that in both cases $J_2=0$ and $J_1=1$. This is a trivial consequence, because the circle is blocked from moving, if the 
particles of the circle stop rotating.     
 
The two invariants (\ref{70.1}) and (\ref{70.2}) were deduced for negligible velocities. 
In the general case the two antisymmetric matrices $C+B$ and $C-B$ should be 
complexified as $C+B+iT$ and $C-B+iT$, and hence we have two complex invariants 
$$I_1=[(d\xi_1,d\xi_2,d\xi_3)+(d\eta_1,d\eta_2,d\eta_3)+ic(d\theta_1,d\theta_2,d\theta_3)]^2,$$
and 
$$I_2=[(d\xi_1,d\xi_2,d\xi_3)-(d\eta_1,d\eta_2,d\eta_3)+ic(d\theta_1,d\theta_2,d\theta_3)]^2.$$
We shall consider two cases: (i) the case of simultaneity ($\Delta t=0$) and (ii) the case of non simultaneity ($\Delta t\neq 0$).  
In the case of simultaneity, having in mind the comments in subsection 2.3, 
analogously to (\ref{70.1}) and (\ref{70.2}) $I_1$ and $I_2$ lead to the following four real invariants: 
\begin{equation}
J_1=d\xi_1^2+d\xi_2^2+d\xi_3^2+d\eta_1^2+d\eta_2^2+d\eta_3^2-
c^2d\theta_1^2-c^2d\theta_2^2-c^2d\theta_3^2,\label{*}
\end{equation}
\begin{equation}
J_2=d\xi_1\cdot d\eta_1+d\xi_2\cdot d\eta_2+d\xi_3\cdot d\eta_3,\label{J_2}
\end{equation}
\begin{equation}
J_3=d\xi_1\cdot d\theta_1+d\xi_2\cdot d\theta_2+d\xi_3\cdot d\theta_3,
\end{equation}
\begin{equation}
J_4=d\eta_1\cdot d\theta_1+d\eta_2\cdot d\theta_2+d\eta_3\cdot d\theta_3. 
\end{equation}

Analogously to (\ref{2.11}) we a priori assume that  
\begin{equation}
(d\theta_x,d\theta_y,d\theta_z)=\vec{l} +\vec{s}\Delta t,\label{deltat}
\end{equation}
where
$$\vec{l}=\frac{1}{c}\Bigl (\frac{d\eta_x}{ds},\frac{d\eta_y}{ds},\frac{d\eta_z}{ds}\Bigr )
\times (d\xi_x,d\xi_y,d\xi_z),$$
$\vec{s}$ is a unit vector which is orthogonal to $\vec{l}$ and $ds=vdt$. 
Hence in the case of simultaneity $\Delta t=0$ we notice that $J_3=J_4=0$. 
The conditions $J_3=J_4=0$ mean that all points of a body which is in rest have the same temporal coordinates 
and also the spatial rotation of this body does not change the temporal coordinates of its points. 
We note that the invariant $J_2$ from (\ref{70.2}) remains unchanged, while the invariant (\ref{70.1}) differs from (\ref{*}) 
by the term 
\begin{equation}
-c^2(d\theta_x^2+d\theta_y^2+d\theta_z^2)
=-\frac{1}{ds^2}
\vert (d\eta_x,d\eta_y,d\eta_z)\times (d\xi_x,d\xi_y,d\xi_z)\vert ^2.\label{**}
\end{equation}
In all of the previous examples about the invariance of (\ref{70.1}) and (\ref{70.2}) the term (\ref{**}) does not change the invariant
(\ref{70.1}), i.e. (\ref{*}) is also an invariant.    

Note that if we have constraints for the spatial rotations (analogous to the section 5), 
and also we have constraint for the spatial displacement, 
i.e. spin velocity, such that the spin velocity is not permitted, then according to the invariant $J_1$ there will appear a temporal displacement, 
which is analogous to the spatial displacement.  
This displacement in time can be called {\em spin time}. 
 
Now let us consider the case of non simultaneity ($\Delta t\neq 0$). We know from subsection 2.3 that 
$Re[(\Delta x)^2+(\Delta y)^2+(\Delta z)^2]$ is an invariant, while $Im[(\Delta x)^2+(\Delta y)^2+(\Delta z)^2]$ is not an invariant.  
In our case (ii) it means that $J_3$ and $J_4$ are not invariants. The invariant $J_2$ from (\ref{J_2}) remains unchanged. The invariant 
$J_1$ from (\ref{*}) takes now the following form 
$$
J_1=(d\vec{\xi})^2+(d\vec{\eta})^2-c^2(d\vec{\theta})^2=
(d\vec{\xi})^2+(d\vec{\eta})^2-
\frac{(d\vec{\xi})^2(d\vec{\eta})^2}{ds^2}(1-\cos^2\varphi)-c^2dt^2,
$$ 
where $\varphi$ is the angle between $\vec{\xi}$ and $\vec{\eta}$. Using that
$$(d\vec{\xi})^2(d\vec{\eta})^2\cos^2\varphi =(d\vec{\xi}\cdot d\vec{\eta})^2=J_2^2,$$ 
we obtain 
$$
J_1=(d\vec{\xi})^2+(d\vec{\eta})^2-
\frac{(d\vec{\xi})^2(d\vec{\eta})^2}{ds^2}-c^2dt^2+\frac{J_2^2}{ds^2},$$
\begin{equation}
J_1=-c^2dt^2\Bigl [1-\Bigl (\frac{d\vec{\eta}}{cdt}\Bigr )^2
-\frac{J_2^2}{c^2dt^2ds^2}\Bigr ].\label{inv}
\end{equation}
Hence the following term 
$$\sqrt{-\frac{J_1}{c^2}}=dt\sqrt{
1-\Bigl (\frac{d\vec{\eta}}{cdt}\Bigr )^2-\frac{J_2^2}{c^2dt^2ds^2}}$$
is also an invariant, 
which is indeed the quantity $d\tau$, where $\tau$ is the proper time of the moving body. 
Since we denote the torsion by $\tau$, the proper time will be denoted further by $t_p$.  
A precise determination of the invariant $J_2$ is given at the end of section 8. Consequently, 
if the considered particle/body moves with a negligible torsion and it has slow variation of the spinning axis, then $J_2$ can be 
neglected and hence 
\begin{equation}
dt_p=dt\sqrt{1-\Bigl (\frac{d\vec{\eta}}{cdt}\Bigr )^2},
\label{tp}
\end{equation}
which is the same as in the Special Relativity. 

At the end we deduce the formula for the energy obtained by a spin motion. 
If we neglect the friction, 
the energy obtained via spin motion from zero up to $\vec{V}$ sometimes can be obtained if  
we integrate the spin acceleration, i.e.  
\begin{equation}
E=\int m\frac{d\vec{V}}{dt}d\vec{r} =\int md\vec{V}\cdot \vec{V}=\frac{1}{2}mV^2.\label{7.4}
\end{equation}
For example, if a spinning body on a horizontal plane in a gravitational field moves on a circle, we will see later that 
this circular motion is indeed a spin motion with spin velocity $V$, and the corresponding kinetic energy now is $\frac{1}{2}mV^2$. 

The spin kinetic energy together with the classical kinetic energy   
$\frac{1}{2}mv^2$ obtained by an inertial velocity $v$ gives the total kinetic energy  
$\frac{1}{2}mv^2+\frac{1}{2}mV^2$. While the ordinary kinetic energy depends on the choice of the 
inertial system, the spin kinetic energy is the same in all inertial systems, because $V$ is the same.   
Let us assume that a non-spinning body with inertial velocity $u$ and 
a spinning body with inertial velocity $v$ and spin velocity $V$, such that $u=v+V$, are moving in a gravitational field.   
Although the observed initial velocities $u$ and $v+V$ are equal, their kinetic energies $\frac{1}{2}mu^2$ and 
$\frac{1}{2}mv^2+\frac{1}{2}mV^2$ are different and their trajectories will be different if $\vec{V}$ is not constant.     

\setcounter{equation}{0} 
\section{Spinning bodies in gravitational field}
\label{sec:7}

The right side of (\ref{6.3}) can easily be written using the radius vector $\vec{r}_i$ of the $i$-th particle (body). 
It leads to the following general formula for the spin velocity 
\begin{equation}
\vec{V}=\Bigl [\sum_i\frac{\vert\vec{r}'_i\vert^4(\vec{r}'_i,\vec{r}''_i,\vec{r}'''_i)}
{\vert\vec{r}'_i\times \vec{r}''_i\vert^4}
(\vec{r}'_i\times \vec{r}''_i)m_i\Bigr ]\frac{1}{\sum_jm_j},\label{8.1}
\end{equation}
where we assume that $\sum_im_i\vec{r}'_i=\vec{0}$, because the barycentre of the bodies should rest at the considered instant. 
If the condition $\sum_im_i\vec{r}'_i=\vec{0}$ is not satisfied, then the following more general formula holds 
\begin{equation}
\vec{V}=\Bigl [\sum_i\frac{\vert\vec{r}'_i-\vec{u}\vert^4(\vec{r}'_i-\vec{u},\vec{r}''_i,\vec{r}'''_i)}
{\vert(\vec{r}'_i-\vec{u})\times \vec{r}''_i\vert^4}
((\vec{r}'_i-\vec{u})\times \vec{r}''_i)m_i\Bigr ]\frac{1}{M}\label{8.2}
\end{equation}
where $\vec{u}=\sum_i\vec{r}'_i\frac{m_i}{\sum_jm_j}$ is the velocity of the barycentre and $M=\sum_jm_j$. 

Now let us consider the case when the body  
is inside a gravitational field with acceleration $\vec{g}$. 
The notion of a gravitational field is only symbolic, because we use only a non-inertial acceleration $\vec{g}$ with respect to any 
coordinate system. Alternatively, one can consider charged bodies in electromagnetic field. 

We shall consider two separate cases, (a) when the body is 
in free fall motion in the gravitational field, and (b) when the spinning body is put on a horizontal plane which is orthogonal to the 
vector $\vec{g}$. 

(a) Assume that the body is in free fall motion in the gravitational field. 
Axiomatically there are two opposite assumptions (as axioms) which are not in contradiction with the previous consideration: 
We may assume that the previous formulas for the case of absence of gravitational acceleration will remain unchanged, or 
the formula for the spin velocity should be modified such that $\vec{r}_i''$ is replaced by $\vec{r}_i''-\vec{g}$. 
The first assumption is in accordance with the Equivalence Principle, while the second assumption comes in a deep contradiction 
with the Equivalence Principle. For this reason we will accept the first assumption. In order to consider its relation with the  
Equivalence Principle we should consider two cases. If the spinning body is axially symmetric and it is spinning around its axis, then 
the spin velocity practically disappears, specially if the angular velocity $w$ is large. But if the spinning axis is not axially 
symmetric, or it is axially symmetric but spins around another axis, then according to the Example 1 in section 5 in this non-symmetric case Newton's third law can be violated and also the Equivalence Principle will be violated. In order for this to occur practically, 
we should change the spinning axis, because it does not change without perturbation or torques. But such an experiment about the 
Equivalence Principle is not mentioned in the literature. Moreover, the air close to the Earth also makes 
perturbations and any 
such phenomena about the spin velocity in case of free fall motion is masked.  

(b) Now, let us consider the case when the spinning body is put on a plane which is orthogonal to the 
vector of acceleration $\vec{g}$. 
This is a more important case and we will devote much more attention to it. 

To the end of this section let us consider a general method for determining the spin velocity for a circle with radius $r$,
which is a subset of an axially symmetric body. 
Without loss of generality we assume that the barycentre is at the coordinate origin, that 
at the initial instant the spin axis is given by $\vec{b}^*=(0,0,1)$ and the circle is given by 
$\{ (r\cos\alpha ,r\sin\alpha ,h): \alpha \in [-\pi ,\pi ]\}$, where $h$ is a constant.   
We denote the spinning vector by $\vec{b}^*$
in order to distinguish from the binormal vectors $\vec{b}$.   
Then an arbitrary trajectory obtained by spinning around the axis of rotation has the form 
\begin{equation}
\vec{r}(t)^T=M(t)\cdot \left [\matrix{r\cos \alpha \cr r\sin \alpha \cr h\cr}\right ],\label{curve}
\end{equation}
where $M(t)$ is an arbitrary orthogonal matrix with $M(0)=I$, and $\alpha$ is arbitrary parameter. 
In the last step in the process of averaging of the 
spin velocity we should integrate by the parameter $\alpha \in [-\pi ,\pi ]$ and then divide by $2\pi$. Let us develop 
the orthogonal matrix $M$ in the Taylor form. In the first approximation we have 
$M=I+\left [\matrix{ 0 & -w & 0\cr w & 0 & 0\cr 0 & 0 & 0\cr}\right ] t$, where $w$ is the angular velocity, while 
in the next two steps up to $t^3$, the condition of orthogonality of $M$ leads to the following form 
$$M=I+\left [\matrix{ 0 & -w & 0\cr w & 0 & 0\cr 0 & 0 & 0\cr}\right ] t+
\left [\matrix{ -w^2 & -p_z & p_y\cr p_z & -w^2 & -p_x\cr -p_y & p_x & 0\cr }\right ]\frac{t^2}{2!}+$$
\begin{equation}
+\left [\matrix{ -3wp_z & w^3-q_z & q_y+\frac{3}{2}wp_x \cr 
-w^3+q_z & -3wp_z & -q_x+\frac{3}{2}wp_y\cr 
-q_y+\frac{3}{2}wp_x & q_x+\frac{3}{2}wp_y & 0\cr }\right ]\frac{t^3}{3!},
\end{equation}
where $(p_x,p_y,p_z)=d\vec{w}/dt$ and $(q_x,q_y,q_z)=d^2\vec{w}/dt^2$. Hence at instant $t=0$ we obtain 
\begin{equation}
\vec{r}'=(-w\sin \alpha ,w\cos \alpha ,0)r,\label{r'}
\end{equation}
$$\vec{r}''-\vec{g}=
(-rw^2\cos \alpha -rp_z\sin \alpha +hp_y-g_x,
-rw^2\sin \alpha +rp_z\cos \alpha -hp_x-g_y,$$
\begin{equation}
-rp_y\cos \alpha +rp_x\sin \alpha -g_z),\label{r''}
\end{equation}
$$\vec{r}'''-\vec{g}'=(-3rwp_z\cos \alpha +rw^3\sin \alpha -q_zr\sin \alpha +h(q_y+\frac{3}{2}wp_x)-{g_x}',$$
$$-3rwp_z\sin \alpha -rw^3\cos \alpha +q_zr\cos \alpha +h(-q_x+\frac{3}{2}wp_y)-{g_y}',$$
\begin{equation}
(-q_y+\frac{3}{2}wp_x)r\cos \alpha + (q_x+\frac{3}{2}wp_y)r\sin \alpha )-{g_z}'),\label{r'''}
\end{equation}

\begin{equation}
\vec{b}=\frac{\vec{r}'\times (\vec{r}''-\vec{g})}{\vert \vec{r}'\times (\vec{r}''-\vec{g})\vert }=
\frac{(A\cos \alpha ,A\sin \alpha ,C)}
{\sqrt{C^2+A^2}},\label{vector-b}
\end{equation}
where
\begin{equation}
A=-rp_y\cos\alpha +rp_x\sin \alpha -g_z,\label{A}
\end{equation}
\begin{equation}
C=rw^2+(g_y+hp_x)\sin \alpha +(g_x-hp_y)\cos \alpha .\label{C}
\end{equation}

The binormal vector $\vec{b}$ depends on the parameter $\alpha$ and does not coincide with the 
unit vector $\vec{b}^*=(0,0,1)$. This shows that the final formula for the spin velocity 
will be very complicated. 
So we assume below that $w$ is sufficiently larger than $\vert \frac{d\vec{b}^*}{dt}\vert$. 

Notice that in (\ref{6.1}) we neglected the term $-(k-k_0)r\vec{t}ds$ from the spin displacement $d\vec{L}$. 
Let us study its influence to the spin velocity and use the same special choice of the coordinate system as previously. 
Notice that $k=k(\alpha )$ is the curvature which corresponds to the trajectory for $\alpha$, $r$ is its radius of curvature, i.e. 
$r=1/k(\alpha )$ and $k_0$ in our case is $1/r$, where $r$ is the radius of the circle. So the averaging for $\alpha \in [-\pi ,\pi ]$
leads to 
$$\langle \frac{d\vec{L}}{ds}\rangle =\langle -\vec{t}+\frac{1}{r}\frac{\vec{t}}{k(\alpha )}\rangle =
\frac{1}{r}\langle \frac{\vec{t}}{k(\alpha )}\rangle .$$
Using that $ds=rwdt$ and $k(\alpha )=\frac{\vert \vec{r}'\times (\vec{r}''-\vec{g})\vert}{r^3w^3}$, where 
$\vec{r}'$ and $\vec{r}''$ were previously given, for the spin velocity we obtain 
$$\vec{V}=\langle \frac{d\vec{L}}{dt}\rangle =w\langle \frac{\vec{t}}{k(\alpha )}\rangle 
=w\langle \frac{r^3w^3\vec{t}}{\vert \vec{r}'\times (\vec{r}''-\vec{g})\vert}\rangle =
r^3w^4\langle \frac{\vec{t}}{\vert \vec{r}'\times \vec{r}''-\vec{r}'\times \vec{g}\vert}\rangle .$$
If there is no gravitation, i.e. $\vec{g}=0$, and if $h=0$, then using the power expansion 
and then using that $\int_{-\pi}^{\pi}\sin^n\alpha \cos^{2m+1-n}\alpha d\alpha =0$ 
for positive integers $n$ and $m$ and $2m+1\ge n$, one can prove that 
$\langle \frac{\vec{t}}{\vert \vec{r}'\times \vec{r}''\vert}\rangle =0$, and hence the spin velocity is 0. 
This is the reason why we neglected the 
term $-(k-k_0)r\vec{t}ds$ in (\ref{6.1}). In order to get formulas for the general case for $h\neq 0$ and $\vec{g}\neq 0$ we will 
consider two special cases and then we will combine them. 

First let us assume that $\vec{g}=0$, but $h\neq 0$. Then according to the previous notations, 
$$\vert \vec{r}'\times \vec{r}''\vert =
r^2w^3\sqrt{\Bigl (\frac{C}{rw^2}\Bigr )^2+(-\frac{p_y}{w^2}\cos \alpha +\frac{p_x}{w^2}\sin \alpha )^2}.$$          
Using that the terms $\sin^2\alpha$, $\cos^2\alpha$ and $\sin \alpha\cos \alpha$ have minor influence in the process of averaging, 
we can assume that 
$$\vert \vec{r}'\times \vec{r}''\vert \approx r^2w^3\Bigl (1+\frac{h}{rw^2}p_x\sin\alpha -\frac{h}{rw^2}p_y\cos\alpha \Bigr ),$$
$$\frac{1}{\vert \vec{r}'\times \vec{r}''\vert} \approx r^{-2}w^{-3}\Bigl (1-\frac{h}{rw^2}p_x\sin\alpha +\frac{h}{rw^2}p_y\cos\alpha \Bigr ).$$
Hence it is easy to obtain that 
$$\vec{V}=r^3w^4\langle \frac{\vec{t}}{\vert \vec{r}'\times \vec{r}''-\vec{r}'\times \vec{g}\vert}\rangle =$$
$$=rw\langle (-\sin \alpha ,\cos \alpha ,0)\cdot 
\Bigl (1-\frac{h}{rw^2}p_x\sin\alpha +\frac{h}{rw^2}p_y\cos\alpha \Bigr )\rangle =$$
$$=rw\frac{h}{rw^2}\bigl (\frac{p_x}{2},\frac{p_y}{2},0)=\frac{h}{2w}(p_x,p_y,0)=
\frac{h}{2w}\Bigl [\frac{d\vec{w}}{dt}-\Bigl (\vec{b}^*\cdot \frac{d\vec{w}}{dt}\Bigr )\vec{b}^*\Bigr ].$$ 
If we use that $\vec{w}=w\vec{b}^*$, where $\vert \vec{b}^*\vert =1$, we obtain that 
$$\vec{V}=\frac{h}{2}\frac{d\vec{b}^*}{dt}.$$

Now assume that $h=0$, but $\vec{g}\neq 0$. The spin velocity is complicated and so we 
assume that $w$ is sufficiently large. Then 
analogously to the previous case one can easily verify that 
$$\vec{V}=\frac{1}{2w}\Bigl (\vec{g}\times \vec{b}^*-\frac{1}{w^2}\frac{d\vec{b}^*}{dt}(\vec{g}\cdot \vec{b}^*)\Bigr ).$$

Now if we combine the last two formulas, we obtain the following formula 
\begin{equation}
\vec{V}=\frac{1}{2w}\Bigl (\vec{g}\times \vec{b}^*-\frac{1}{w^2}\frac{d\vec{b}^*}{dt}(\vec{g}\cdot \vec{b}^*)\Bigr )
+\frac{h}{2}\frac{d\vec{b}^*}{dt}\label{V-curvature},  
\end{equation}
which is a consequence of the term $-(k-k_0)r\vec{t}ds$ in (\ref{6.1}). 
Remember that this spin velocity corresponds to the spin velocity of the circle 
$\{ (r\cos \alpha ,r\sin\alpha ,h)\}$. When we sum over all such circles and use that coordinate origin coinciding with the 
barycentre of the axially symmetric spinning body, we obtain that $\int \frac{h}{2}\frac{d\vec{b}^*}{dt}dm=
(\int \frac{h}{2}dm)\frac{d\vec{b}^*}{dt}=0$, 
such that the summand $\frac{h}{2}\frac{d\vec{b}^*}{dt}$ has no role in this case. So, for this reason and for the case of simplicity 
we will omit the terms which contain $h$, i.e. we will assume that $h=0$. 

\setcounter{equation}{0} 
\section{Exact formula for the spin velocity}
\label{sec:8}

We studied this theory step by step. 
In section 5 we mentioned that 
the differential of the matrix $A(t)$ is given by 
$$\left [\begin{array}{ccc} 0 & k_0 & 0 \\ -k_0 & 0 & 0 \\ 0 & 0 & 0\end{array}\right ]ds,$$
where $k_0=\frac{wdt}{ds}$, with respect to a moving trihedron which is close to $(\vec{t},\vec{n},\vec{b})$. 

In this section we will find the exact spin velocity of a single point, where this approximation will vanish. It will be just the 
exact expression for the spin velocity. 

Let us use the notations in the section 7. 
Then we obtain successively 
$$\vec{r}'=rw(-\sin \alpha ,\cos \alpha ,0),\quad \vec{t}=(-\sin \alpha ,\cos \alpha ,0),$$
$$\vec{b}=\Bigl (\frac{A\cos \alpha}{\sqrt{C^2+A^2}},\frac{A\sin \alpha}{\sqrt{C^2+A^2}},\frac{C}{\sqrt{C^2+A^2}}\Bigr ),$$ 
$$\vec{n}=\vec{b}\times \vec{t}=\Bigl (\frac{-C\cos \alpha }{\sqrt{C^2+A^2}},\frac{-C\sin \alpha }{\sqrt{C^2+A^2}},
\frac{A}{\sqrt{C^2+A^2}}\Bigr ),$$ 
where $A$ and $C$ are defined by (\ref{A}) and (\ref{C}) respectively.  

Let us introduce vectors $\vec{t}^*$ and $\vec{n}^*$ by 
$$\vec{t}^*=(-\sin \alpha ,\cos \alpha ,0),\quad \vec{n}^*=(-\cos \alpha ,-\sin \alpha ,0),$$ 
and introduce an angle $\psi$ such that $\cos \psi =\frac{C}{\sqrt{C^2+A^2}}$ and $\sin \psi =\frac{A}{\sqrt{C^2+A^2}}$. 
Now the connection between the trihedrons $(\vec{t}^*,\vec{n}^*,\vec{b}^*)$ and $(\vec{t},\vec{n},\vec{b})$ is given by 
$$\vec{t}=\vec{t}^*,\; \vec{n}=\cos \psi \vec{n}^*+\sin \psi \vec{b}^*, \; \vec{b}=\cos \psi \vec{b}^*-\sin \psi \vec{n}^*.$$

If there are no constraints, according to the Frenet equations we almost know that   
the differential of the spatial rotation is given by 
$$
-\tau \vec{t}ds -k \vec{b}ds,
$$
i.e. 
\begin{equation}
-\tau \vec{t}^*ds +k\sin\psi \vec{n}^*ds-k\cos \psi \vec{b}^*ds.\label{9.0}
\end{equation}

On the other side, the differential of the admitted spatial rotation with respect to the trihedron 
$(\vec{t}^*,\vec{n}^*,\vec{b}^*)$ is given by 
$$\left [\matrix{ 0 & 1/r & 0 \cr -1/r & 0 & 0 \cr 0 & 0 & 0\cr }\right ] ds,$$
which is spatial rotation for the vector $-1/r\vec{b}^*ds$. 
The spin velocity can be obtained by multiplication from the right of the subtraction of these two vectors of angular rotations 
with the vector $-\frac{1}{k}\vec{n}=-\frac{1}{k}(\cos \psi \vec{n}^*+\sin \psi \vec{b}^*)$. 
Using also that $ds=rwdt$, 
the spin velocity of the considered point after some transformations can be written in the following form  
\begin{equation}
\vec{V}=w\Bigl (\frac{\cos \psi}{k}-r\Bigr )\vec{t}+\Bigl (\frac{\tau}{k}wr\Bigr )\vec{b}.
\label{9.1}
\end{equation}
The formula (\ref{9.1}) is the {\em exact formula for the spin velocity}.

If we want to average this spin velocity around the circle with radius $r$ or the spinning body, 
the vectors $\vec{t}$ and $\vec{b}$ should be replaced respectively by $(-\sin \alpha ,\cos \alpha ,0)$ 
and $\sin \psi (\cos \alpha ,\sin \alpha ,0)+\cos \psi (0,0,1)$, $k(\alpha )$ and $\tau (\alpha)$ should be calculated for the curve (\ref{curve}), 
while 
$r$ and $w$ are invariants. The required spin velocity should be written as a function of 
$\vec{b}^*$ and $w$ and their derivatives. The vector $\vec{b}^*$ is the unit vector of the spinning axis, 
which in the chosen coordinate system and chosen instant is $(0,0,1)$.  

Now let us start with the exact spin velocity (\ref{9.1}) in order to check whether $J_1$, $J_2$, $J_3$ and $J_4$ are really invariants
in the case of simultaneity $(\Delta t=0)$, analogously to Example 1 from section 6. 
Using the notations from the section 6 we obtain the following. 
If there are no constraints, then using the angular velocity (\ref{9.0}) we obtain 
\begin{equation}
d\vec{\eta}=\vec{t}^*ds,\label{9.a}
\end{equation}
\begin{equation}
d\vec{\xi}=(\tau \vec{t}^*-k\sin \psi \vec{n}^* + k\cos \psi \vec{b}^*)\frac{ds}{k},\label{9.b}
\end{equation}
while in the case of constraints using the spin velocity (\ref{9.1}) we obtain 
$$
d\vec{\eta}_{con.}=\vec{t}ds+\vec{V}dt=
\vec{t}ds+ \Bigl [w\Bigl (\frac{\cos \psi}{k}-r\Bigr )\vec{t}+\Bigl (\frac{\tau}{k}wr\Bigr )\vec{b}\Bigr ]dt=
$$
\begin{equation}
=\Bigl [\frac{w\cos \psi}{k}\vec{t}+\Bigl (\frac{\tau}{k}wr\Bigr )\vec{b}\Bigr ]dt
=\frac{\cos \psi}{rk} \vec{t}ds +\frac{\tau}{k}\vec{b}ds,\label{9.c}
\end{equation}
\begin{equation}
d\vec{\xi}_{con.}=\vec{b}ds.\label{9.d} 
\end{equation}
The explanation of formula (\ref{9.d}) will be given in the following section.  
Now let us calculate the values $J_1$, $J_2$, $J_3$ and $J_4$ in both cases if there are no constraints and with constraints. 
According to (\ref{9.a}), (\ref{9.b}) and (\ref{**}) we obtain 
$$J_1=2ds^2+\Bigl (\frac{\tau ds}{k}\Bigr )^2-c^2(d\theta _1^2+d\theta _2^2+d\theta _3^2)=
ds^2+\Bigl (\frac{\tau ds}{k}\Bigr )^2,$$
$$J_2=\frac{\tau}{k}ds^2,$$
while according to (\ref{9.c}), (\ref{9.d}) and (\ref{**}) in the case of constraints we obtain 
$$(J_1)_{con.}=ds^2+\frac{\cos^2\psi}{k^2r^2}ds^2+\Bigl (\frac{\tau ds}{k}\Bigr )^2-
c^2[(d\theta _1)_{con.}^2+(d\theta _2)_{con.}^2+(d\theta _3)_{con.}^2]=$$
$$=ds^2+\Bigl (\frac{\tau ds}{k}\Bigr )^2,$$ 
$$(J_2)_{con.}=\frac{\tau}{k}ds^2.$$
We notice that $J_1=(J_1)_{con.}$, $J_2=(J_2)_{con.}$ and also $J_3=(J_3)_{con.}=0$ and $J_4=(J_4)_{con.}=0$ are trivially satisfied. 
Note also that the vectors $(d\theta _1,d\theta _2,d\theta _3)$ and $((d\theta _1)_{con.},(d\theta _2)_{con.},(d\theta _3)_{con.})$ 
are collinear, because both of them are collinear with $\vec{n}$.  

{\bf Remark 1.} In order for the proper time $t_p$ to be a non-negative real number, after calculation of the invariant $J_2$ we obtain the following condition 
\begin{equation}
v^2+\frac{\tau^2}{k^2}v^2 \le c^2.
\end{equation} 
On the other hand, according to (\ref{9.c})
$$\Bigl\vert \frac{d\vec{\eta}_{con.}}{dt}\Bigr\vert ^2 = \Bigl (\frac{\cos\psi }{rk}v\Bigr )^2+
\frac{\tau^2}{k^2}v^2$$
can be larger than $c^2$, because $\Bigl (\frac{\cos\psi }{rk}\Bigr )^2$ can be larger than 1. This means that 
the observed velocity $\Bigl\vert \frac{d\vec{\eta}_{con.}}{dt}\Bigr\vert$ can be larger than $c$. 
      
{\bf Example 1.} Let us consider a circle with radius $r$ and a constant angular velocity $\vec{w}$, i.e. $w$ is a constant and 
$d\vec{b}^*/dt=0$, and assume that $\vec{b}^*$ is orthogonal to $\vec{g}$. If $v=rw<c$, then the proper time of the points of the circle 
is $dt_p=dt\sqrt{1-r^2w^2/c^2}$, because $J_2=0$ since $\tau =0$ in this case. 
Further, according to (\ref{9.4}) and (\ref{V-curvature1}) the spin velocity in section 10 is estimated by  
$$\vec{V}=\frac{r^2w^3(r^2w^4-\frac{1}{2}g^2)}{2(r^2w^4+\frac{1}{2}g^2)^2}(\vec{g}\times \vec{b}^*).$$
Hence if $\vec{v}=\vec{w}\times\vec{r}$ is collinear with $\vec{g}\times \vec{b}^*$ and $v\approx c$, the observed velocity 
$\vec{v}+\vec{V}$ can be larger than $c$. 
    
\setcounter{equation}{0} 
\section{On the precession of the spinning axis}
\label{sec:9}

Let us return to the equation (\ref{9.d}).  
Analogously as the constrained angular velocity leads to the spin displacement and hence 
spin velocity, it also tends to change the spinning axis, such that the whole body would be additionally rotated. 
This influence maps the previous spinning axis 
determined by the vector $\vec{b}^*$ into the binormal vector $\vec{b}$ of the trajectory of the considered point, 
because each point tends to rotate in self osculating plane free of torsion.  
We will take into account these 
changes of the spinning axes of all points, and it will lead to the joint spinning axis. 
These precessions of the spinning axes are analogous to the spin
displacements of the points. The spin displacement as well as 
the rotations do not realize completely, and so their constraints lead to the change of temporal coordinates.

Let us calculate now the precession of the spinning axis.  
The rotation which maps the vector $\vec{b}^*$ into the vector 
$\vec{b}$ is given by the 3-vector $\psi\vec{t}$. First we assume that there are not any torques, which means that 
$(p_x,p_y,p_z)=0$ in the term for $\psi$. Later the precessions caused by any torques or tidal forces should be added.  
The considered effect disappears if $\vec{g}=0$,  
and it also disappears in the case of free fall motion, assuming that $\vec{g}$ is everywhere constant.  
The influence to the precession of the Earth's spinning axis will be considered later.  

Since 
$$\langle \psi \vec{t}\rangle =\langle\tan^{-1}\frac{A}{C}(-\sin \alpha ,\cos \alpha ,0)\rangle $$
we obtain the following shift for $p_x$ and $p_y$
\begin{equation}
-\frac{1}{2\pi}\int_{-\pi}^{\pi} \tan^{-1}\frac{A}{C}\sin \alpha d\alpha =\frac{\Delta p_x}{w^2}, \label{inteq1}
\end{equation} 
\begin{equation}
\frac{1}{2\pi}\int_{-\pi}^{\pi} \tan^{-1}\frac{A}{C}\cos \alpha d\alpha =\frac{\Delta p_y}{w^2}, \label{inteq2}
\end{equation}
where $A$ and $C$ are given by (\ref{A}) and (\ref{C}). The components $p_x$ and $p_y$, which appear in the terms $A$ and $C$, 
arise from some possible torques which change the spinning axis. These components do not have anything joint with the required components $\Delta p_x$ 
and $\Delta p_y$. In the case of weak gravitational field the influence of $(p_x,p_y,0)$ and $\vec{g}$ can be superposed and 
so without loss of generality we assumed that $p_x=p_y=0$, and now let us denote $\Delta p_x$ and $\Delta p_y$ by $p_x$ 
and $p_y$. 

If we multiply (\ref{inteq1}) by $-p_y$ and add to the equation (\ref{inteq2}) multiplied by $p_x$ we obtain the following equation 
\begin{equation}
\int_{-\pi}^{\pi} \tan^{-1}\frac{A}{C}(p_x\cos \alpha +p_y\sin\alpha )d\alpha = 0.\label{inteq3}
\end{equation}

We will prove now that $p_x:p_y=(-g_y):g_x$ satisfies the above system,  
which means that $d\vec{b}^*/dt$ is parallel to $\vec{b}^*\times \vec{g}=(-g_y,g_x,0)$.  
If we replace $p_x=-ag_y$ and $p_y=ag_x$ into (\ref{inteq3}), we obtain  
$$p_x\cos \alpha +p_y\sin \alpha =-a( rw^2+g_y\sin \alpha +g_x\cos \alpha \bigr) ',$$ 
$$\frac{A}{C}=\frac{-g_z}{rw^2+g_y\sin \alpha +g_x\cos \alpha},$$
and hence 
$$\tan^{-1}\frac{A}{C}=F(rw^2+g_y\sin \alpha +g_x\cos \alpha ).$$ 
Thus, 
$$\int_{-\pi}^{\pi} \tan^{-1}\frac{A}{C}(p_x\cos \alpha +p_y\sin\alpha )d\alpha = $$
$$=(-a)\int_{-\pi}^{\pi}F(rw^2+g_y\sin \alpha +g_x\cos \alpha \bigr )
d(rw^2+g_y\sin \alpha +g_x\cos \alpha )=$$
$$=(-a)\Phi (rw^2+g_y\sin \alpha +g_x\cos \alpha \bigr )\Bigl\vert _{-\pi}^{\pi}=0$$
and the proof is finished. 

Let us determine $d\vec{b}^*/dt$. 
If we replace $p_x=-ag_y$ and $p_y=ag_x$ into (\ref{inteq1}) or (\ref{inteq2}), 
we can solve that equation by $a$ and then  
$(p_x,p_y,0)=(-ag_y,ag_x,0) = a(\vec{b}^*\times\vec{g})$ at the considered point. 
Since the value $a$ does not depend on the choice of the coordinate system, assuming that $\vec{b}^*=(0,0,1)$, we may assume that 
$g_y=0$ in the chosen coordinate system and hence the equation (\ref{inteq1}) is identically satisfied, 
while (\ref{inteq2}) becomes 
$$\frac{1}{2\pi}\int_{-\pi}^{\pi} \tan^{-1}\frac{-g_z}{rw^2+g_x\cos \alpha}\cos \alpha d\alpha =\frac{ag_x}{w^2}.$$
Hence the required vector $(p_x,p_y,0)$ is given by 
$$(p_x,p_y,0)=(0,w^2J,0),$$
where 
$$J=\frac{1}{2\pi}\int_{-\pi}^{\pi} \tan^{-1}\frac{-g_z}{rw^2+g_x\cos \alpha}\cos \alpha d\alpha .$$
This formula was deduced under the assumptions $\vec{b}^*=(0,0,1)$ and $g_y=0$. In the general case, the previous formula 
can be extended for arbitrary vectors $\vec{b}^*$ and $\vec{g}$, and hence for the vector $d\vec{b}^*/dt=(p_x,p_y,p_z)/w$ we obtain
in final form 
\begin{equation}
\frac{d\vec{b}^*}{dt}=\frac{\vec{b}^*\times \vec{g}}{\vert\vec{b}^*\times \vec{g}\vert}wJ,\label{db}
\end{equation}
where 
\begin{equation}
J=\frac{1}{2\pi}\int_{-\pi}^{\pi}\tan^{-1}\frac{-(\vec{b}^*\cdot \vec{g})}
{rw^2+\vert \vec{b}^*\times \vec{g}\vert\cos \alpha}\cos \alpha d\alpha .\label{J}
\end{equation}

The formula (\ref{db}) was deduced if the spinning body is a circle with radius $r$, and 
(\ref{db}) is just the average precession caused by an arbitrary chosen point of the circle per one cycle. 
If we want to find a general formula 
for an arbitrary spinning body which is an axially symmetric, we should divide (\ref{db}) by $J$ and then we are permitted 
to integrate over the whole spinning body. Hence we obtain the following general formula 
\begin{equation}
\frac{d\vec{b}^*}{dt}=\frac{\vec{b}^*\times \vec{g}}{\vert\vec{b}^*\times \vec{g}\vert}
Mw \Bigl(\int J^{-1}\; dm \Bigr)^{-1},\label{db1}
\end{equation}
where $J$ is given (\ref{J}) and $M$ is the mass of the spinning body.   

The general formula (\ref{db1}) can be used    
for an arbitrary spinning body on a horizontal plane, instead of a spinning circle. 
In a special case if ${\vert\vec{g}\vert}<<rw^2$, formula (\ref{db1}) reduces to the following simple form 
\begin{equation}
\frac{d\vec{b}^*}{dt}=\frac{(\vec{b}^*\times \vec{g})(\vec{b}^*\cdot \vec{g})M}
{2Iw^3},\label{db2}
\end{equation}
where $I$ is the moment of inertia of the body. 
If we write the vector $\vec{b}^*$ in the following form 
$\vec{b}^*=(\sin\varphi \cos \Omega t,\sin\varphi \sin\Omega t,\cos \varphi )$, 
then for the angular velocity $\Omega$ we obtain 
$\vert\Omega\vert =\frac{g}{\vert\vec{b}^*\times \vec{g}\vert}\bigl\vert\frac{d\vec{b}^*}{dt}\bigr\vert$.  
The approximative formula (\ref{db2}) can be deduced more easily directly from (\ref{inteq1}) and (\ref{inteq2}) if we use 
the condition ${\vert\vec{g}\vert}<<rw^2$. The vector $\frac{d\vec{b}^*}{dt}$ is not always of type 
$\propto \frac{1}{I}$. Indeed, if ${\vert\vec{g}\vert}<rw^2$ instead of (\ref{db2}) the following formula holds
\begin{equation}
\frac{d\vec{b}^*}{dt}=\frac{(\vec{b}^*\times \vec{g})(\vec{b}^*\cdot \vec{g})Mw}
{2Iw^4-\frac{3}{2}M(\vec{b}^*\times \vec{g})^2+\cdots }.\label{db2a}
\end{equation}

The precessions (\ref{db1}) and (\ref{db2}) 
do not include any precession from torques or tidal forces, except the considered gravitational effect.   
Thus this precession is a complementary to the classical formulas for precession. It is obvious according to the following 
example for the precession of the Earth's spinning axis caused by the Sun and Moon. 
Since the Earth is in a free fall motion, the acceleration $\vec{g}$ from (\ref{db2}) should be replaced by 
$\vec{g}-\vec{g}_b$, where $\vec{g}_b$ is the gravitation acceleration at the barycentre of the Earth.  
Hence the formula (\ref{db2}) reduces to  
\begin{equation}
\frac{d\vec{b}^*}{dt}=\frac{\int(\vec{b}^*\times (\vec{g}-\vec{g}_b))(\vec{b}^*\cdot (\vec{g}-\vec{g}_b))dm}
{2Iw^3}.\label{db3}
\end{equation}
Let us assume that at the considered instant the coordinate system is such that the center of the Earth is at the 
origin, $\vec{b}^*=(0,0,1)$ and the radius vector of the Sun is given by $\vec{R}=\vec{n}R$, where  
$\vec{n}=(\cos\gamma \cos \alpha ,\sin \alpha ,\sin \gamma \cos\alpha )$ and $\gamma =23.43928^0$. 

Further we use that 
$$\Delta \vec{g}=\vec{g}-\vec{g}_b=-3\frac{GM_s}{R^3}\vec{n}\Delta R +\frac{GM_s}{R^3}\Delta \vec{R},$$
where $\Delta R=-\Delta \vec{r}\cdot \vec{n}$, $\Delta \vec{R}=-\Delta \vec{r}$,  
$M_s$ is the mass of the Sun and $R$ is the distance between the Earth and Sun. After integration in (\ref{db3}) and 
averaging we obtain 
$$\frac{d\vec{b}^*}{dt}=\frac{\vec{b}^*\times \vec{g}}{\vert\vec{b}^*\times \vec{g}\vert}
\frac{3G^2M_s^2\cos \gamma \sin \gamma}{8w^3R^6}.$$ 
Using that the angular velocity for this precession is $\Omega = \frac{1}{\sin \gamma}\Big\vert \frac{d\vec{b}^*}{dt}\Big\vert$ and 
using that the distance $R$ is not a constant we can obtain the following more precise formula 
\begin{equation}
\Omega_{s}= \frac{3}{8}\frac{G^2M_s^2\cos \gamma}{w^3a^6(1-e^2)^3},\label{Sun} 
\end{equation}
where $e$ is eccentricity and $a$ is the semimajor axis of the Earth's orbit. This leads to the value 
$$\Omega _s= 0.009083''/\hbox{yr}.$$ 

If we consider the influence from the Moon, then a similar formula holds, but we should additionally include the inclination 
$i\approx 5.1567^0$ of the Moon's orbit relative to the ecliptic and we obtain
\begin{equation}
\Omega_{m}= \frac{3}{8}\frac{G^2M_m^2(1-1.5\sin^2 i)^2\cos \gamma}{w^3a^6(1-e^2)^3},\label{Moon} 
\end{equation}
where $e$ is the eccentricity and $a$ is the semimajor axis of the relative Earth-Moon's orbit and $M_m$ is the mass of the Moon. 
This leads to the value 
$$\Omega _m= 0.043055''/\hbox{yr}.$$

Note that according to (\ref{db3}) the precession is not linear, but quadratic with respect to the gravitational acceleration $g$. 
So instead of using separate accelerations toward the Moon and the Sun, it is more correct to use their sum. 
But however, using the formula 
$$\langle (a\sin w_1t +b\sin w_2t)^2\rangle 
\approx \langle (a\sin w_t )^2\rangle +\langle (b\sin w_2t)^2\rangle $$ 
for $w_1 \neq w_2$, on long time intervals it is permitted to consider these two influences separately. 
Their sum
$$\Omega =\Omega _s +\Omega _m = 0.052138''/\hbox{yr}$$
is about $0.1\%$ from the total observed value of about $50.29''/\hbox{yr}$, which is mainly induced by the 
Sun's, Moon's and planetary torques. 

Note also that both of the directions of the precessions (\ref{Sun}) and (\ref{Moon})  
are in the same direction as the trajectory of the Earth around the Sun. 
While the precession caused by the torques is non-zero because of the equatorial bulge of the Earth, 
the precessions (\ref{Sun}) and (\ref{Moon}) remain the same if the Earth was an ideally spherical and uniform body.     
The considered precession in this section does not cause nutation of the Earth's spinning axis and the spinning axis of a body on the horizontal plane.  

\setcounter{equation}{0} 
\section{Spin velocity in a gravitational field and some examples}
\label{sec:10}

After we found the formula for the precession of the spinning axis in a gravitational field, we are able finally to present  
and calculate the spin velocity for a circle with radius $r$. 
Two cases may occur: (i) if the gravitational force is a unique perturbing field and (ii) besides  
the gravitational field there is another external perturbation  
which changes the spinning axis of the considered body. 

(i) In this case 
$d\vec{b}^*/dt$ and hence $(p_x,p_y,p_z)$ is given by (\ref{db1}) and (\ref{db2}).  
Let us start with formula (\ref{9.1}).  
Since $\langle wr\vec{t}\rangle =0$, we should consider the averaging  of the spin velocities  
$\frac{w\cos \psi}{k}\vec{t}$ and $wr\frac{\tau}{k}\vec{b}$. 
We should use the formulas (\ref{r'}), (\ref{r''}) and (\ref{r'''}) for the first three derivatives including the influence from the 
gravitational acceleration. 
Assuming that $w$ is sufficiently larger than $\vert \frac{d\vec{b}^*}{dt}\vert$, we may use that for the $i$-th particle of the body 
$\frac{\vec{r}_i^{'}\times \vec{r}_i^{''}}{\vert \vec{r}_i^{'}\times \vec{r}_i^{''}\vert }\approx \vec{b}^*$.  
It is a problem to average the 
spin velocity for $h\neq 0$. Sometimes $h$ has no influence to the final result, and so we will neglect it in the formulas below.
But sometimes for high order terms, for example 
with $h^2$, it has influence on the spin velocity.   
We will use some other approximations,  
because sometimes the corresponding integrals can only  
numerically be solved. In order to avoid large expressions, 
the vectors $\vec{g}'$ and $\vec{q}=d^2\vec{w}/dt^2$ will be neglected.    

The averaging of $\frac{w\vec{t}}{k}$ led to the spin velocity (\ref{V-curvature}), but now the averaging of    
$\frac{w\cos \psi \vec{t}}{k}$ leads to the following part of the spin velocity 
\begin{equation}
\vec{V}=\frac{r^2w^3}{2U^2}(2r^2w^4-U)(\vec{g}\times \vec{b}^*)-\frac{r^4w^6}{U^2}(\vec{b}^*\cdot \vec{g})
\frac{d\vec{b}^*}{dt},\label{V-curvature1} 
\end{equation}
where 
\begin{equation}
U\approx (rw^2)^2 +(\vec{g}\cdot \vec{b}^*)^2+\frac{1}{2}g^2+\frac{1}{2}r^2w^2\Bigl (\frac{d\vec{b}^*}{dt}\Bigr )^2\label{U}
\end{equation}
is averaging of the term $A^2+C^2$.   

Let us consider now the averaging of $wr\frac{\tau}{k}\vec{b}$.  
We can use (\ref{8.1}) and it is convenient to decompose the required velocity as   
$$
\vec{V}=\lambda{\vec V}_I+\lambda{\vec V}_{II} +\lambda{\vec V}_{III}+\lambda{\vec V}_{IV}, 
$$
where  
$${\vec V}_I= \vec{g}\cdot (\vec{r}'\times \vec{r}''')
\frac{\vert\vec{r}'\vert^4(\vec{r}'\times \vec{r}'')}
{\vert \vec{r}'\times \vec{r}''\vert^4},\quad 
{\vec V}_{II} = -\vec{g}\cdot (\vec{r}'\times \vec{r}''')
\frac{\vert\vec{r}'\vert^4(\vec{r}'\times \vec{g})}
{\vert \vec{r}'\times \vec{r}''\vert^4},
$$
$${\vec V}_{III} = -(\vec{r}',\vec{r}'',\vec{r}''')
\frac{\vert\vec{r}'\vert^4(\vec{r}'\times \vec{g})}
{\vert \vec{r}'\times \vec{r}''\vert^4},\quad 
{\vec V}_{IV} = (\vec{r}',\vec{r}'',\vec{r}''')
\frac{\vert\vec{r}'\vert^4(\vec{r}'\times \vec{r}'')}
{\vert \vec{r}'\times \vec{r}''\vert^4},$$
$$\lambda =\frac{\vert \vec{r}'\times \vec{r}''\vert^4}{\vert \vec{r}'\times (\vec{r}''-\vec{g})\vert^4}. $$
The sum of averaging of these components yields the following spin velocity  
$$\vec{V}= \frac{3\lambda_0}{4w^2}\Bigl (\vec{g}\cdot \frac{d\vec{b}^*}{dt}\Bigr )\vec{b}^*
+3\lambda_0 \frac{w'}{w^3}(\vec{g}\cdot \vec{b}^*)\vec{b}^*+$$
$$+\frac{\lambda_1}{rw^4}
\Bigl [\frac{3}{4}\Bigl (\vec{g}\cdot \frac{d\vec{b}^*}{dt}\Bigr )
+3\frac{w'}{w}(\vec{g}\cdot \vec{b}^*)\Bigr ]
\frac{\vec{g}\times (\vec{g}\times \vec{b}^*)}{\vert \vec{g}\times \vec{b}^*\vert }+
\frac{\lambda_1w'r}{2w^3}
\Bigl (\frac{d\vec{b}^*}{dt}\Bigr )^2
\frac{ \vec{g}-\vec{b}^*(\vec{g}\cdot \vec{b}^*)}{\vert \vec{g}\times \vec{b}^*\vert}+$$
\begin{equation}
+\frac{\lambda_0+\lambda_2}{2w^2}\Bigl [\vec{g}\times 
\Bigl( \vec{b}^*\times \frac{d\vec{b}^*}{dt}\Bigr)\Bigr ] +
\frac{\lambda_2}{w^2}
(\vec{g}\cdot \vec{b}^*)\Bigl (\frac{\vec{g}\times \vec{b}^*}{\vert \vec{g}\times \vec{b}^*\vert}\cdot \frac{d\vec{b}^*}{dt}\Bigr )
\frac{\vec{g}\times \vec{b}^*}{\vert \vec{g}\times \vec{b}^*\vert},\label{9.4}
\end{equation}
where 
\begin{equation}
\lambda_p=\frac{1}{2\pi}\int_{-\pi}^{\pi}\frac{r^4w^8\cos pt\; dt}{(A^2+C^2)^2},\label{lambdap}
\end{equation}
for $p=0,1,2$, and where $A$ and $C$ are given by (\ref{A}) and (\ref{C}). 
In the terms of $A$ and $C$, as well as in the above formulas 
(\ref{9.4}) and (\ref{V-curvature1})  
$d\vec{b}^*/dt=(p_x,p_y,p_z)$ is determined by (\ref{db1}) and (\ref{db2}).  

Now the formula (\ref{9.4}) together with (\ref{V-curvature1}) give the required total spin velocity with some approximations.  
Let us consider some conclusions from (\ref{9.4}) and (\ref{V-curvature1}).  
We present three simple examples and assume in all of them that 
$\vec{g}=(0,0,-g)$, where $g$ is a constant. All these examples can be verified by experiments. 

{\bf Example 1.} Let us assume that the unit vector $\vec{b}^*(t)$ of spinning axis of a spinning circle is given by   
$$\vec{b}^*=(\sin\varphi \cos \Omega t,\sin\varphi \sin\Omega t,\cos \varphi ),$$
where $\Omega$ is a constant, $\varphi\in [0,\pi /2)$ is also a constant and assume also that the angular velocity $w$ 
is sufficiently larger constant than $\Omega$. 
The spin velocity is determined basically by the 
sum of the spin velocities (\ref{9.4}) and (\ref{V-curvature1}). Since $w=const.$ by assumption, 
all terms containing $w'$ will be zero there. 

Notice that in all terms the spin velocity is parallel to $(-\sin\Omega t,\cos \Omega t,0)$, while
its derivative, i.e. the spin acceleration is parallel to $(\cos\Omega t,\sin \Omega t,0)$. Hence we can conclude that 
the spinning body, i.e. its barycentre, moves on a circle with the same angular velocity $\Omega$ as the angular velocity for the 
unit vector $\vec{b}^*$, and moreover, $\vec{V}$ is orthogonal to the unit vector $\vec{b}^*$.
The radius of the circle of motion of the spinning body is equal to 
$R_{circle}=\frac{\vert \vec{V}\vert }{\vert \Omega\vert}$.   

{\bf Example 2.}  
Let us consider a spinning disc, whose mass is distributed on a distance $r$ 
from the center and assume that the vector $\vec{b}^*$ is a constant, while $w$ varies.  
The spin velocity is determined basically by the 
sum of the spin velocities (\ref{9.4}) and (\ref{V-curvature1}), but we will draw more attention to the 
following part of the spin velocity which contains $w'$:  
\begin{equation}
\vec{V}=\frac{3\lambda_0w'}{w^3}(\vec{g}\cdot \vec{b}^*)\vec{b}^* 
+\frac{3w'\lambda_1}{rw^5}(\vec{g}\cdot \vec{b}^*)
\frac{\vec{g}\times (\vec{g}\times \vec{b}^*)}{\vert \vec{g}\times \vec{b}^*\vert }+
\frac{\lambda_1w'r}{2w^3}
\Bigl (\frac{d\vec{b}^*}{dt}\Bigr )^2
\frac{ \vec{g}-\vec{b}^*(\vec{g}\cdot \vec{b}^*)}{\vert \vec{g}\times \vec{b}^*\vert}+\dots 
\label{9.7} 
\end{equation} 

We deduce the following important conclusion from the equality (\ref{9.7}). It is convenient to 
increase the angular velocity of the disc from 0 up to approximately $w=\sqrt {\frac{g}{r}}$.  
Since the spin velocity $\vec {V}$ mainly is proportional to $w'$ according to (\ref{9.7}), we can 
increase the angular velocity slowly, such that the spin velocity is almost negligible. 
So we input an energy $E$ only for the kinetic energy of the rotating disc.    
Further let us quickly decrease this angular velocity up to zero angular velocity ($w=0$). 
In this process we can return the input kinetic energy $E$. Moreover, since $w'\neq 0$ means that the spin velocity $\vec {V}$ is non-zero, 
the derivative of the spin velocity $\vec{V}$  
makes pressure, which is indeed {\em free energy}. 
We can repeat this process periodically as many times as we wish, in order to obtain the required quantity of energy. 
In practice the friction is always non-zero, but we should decrease the angular velocity of the disc very quickly such that 
the free energy is larger than the energy spent for the friction. 

The total free energy is equal to $\frac{M}{2}{\vec V}^2$. If $w$ decreases by $w(t)=\sqrt{\frac{g}{r}}\bigl (1-\frac{t}{t_0}\bigr )$
$(0\le t\le t_0)$,  
then in this case the free energy is of order $M\frac{r^2}{t_0^2}$.  
So, we do not need to return back the input kinetic energy $E=\frac{1}{2}Mr^2w^2$,
because the derived free energy can be sufficiently larger than $E$. 
If we use this free energy such that 
the disc in not permitted to be displaced freely, then there will be displacement in time, i.e. spin time displacement, 
and according to the invariant $J_1$ this will cause the energy of the disc to fall. This loss of the energy of the disc 
is analogous to the loss of energy of the bodies which are placed closer to the center of a massive gravitational body. 
Consequently the mass of the disc will be slightly decreased and the speed of the proper time of the disc will 
be slightly decreased too. So the free energy is simply energy subtracted from the energy of the disc.   

This is not a unique way to derive free energy using only 
mechanics. The free energy can be obtained also in a gravitational or electromagnetic field by 
slow increase of $d\vec{b}^*/dt$ and then quickly decreasing, 
but the previous procedure is the simplest way to show the existence of the free energy. 
The author of this paper has measured the weight of a spinning disc  
when its angular velocity decreases and the angle between $\vec{b}^*$ and the horizontal plane is $45^0$. 
When the angular velocity was close to a value $w_0$, the weight was slightly increased. 
Indeed, for $w \rightarrow 0$ and $w\rightarrow \infty$ the spin velocity tends to 0. So, there exists $w_0$ such that 
$\vec{V}(w)$ has a maximum.  
When the disc slowly decreased 
its angular velocity because of friction, the variation of the weight was very small. But when the angular velocity was 
strongly manually decreased, 
the change of the weight was much larger, because the free energy is proportional to $(w')^2$.   
However, the increasing of the weight of the disc confirms the existence of free energy. 

(ii) Assume that besides  
the gravitational field and its influence 
there are also external perturbations or torques 
which change the spinning axis of the considered body. 
Besides previously determined vector $(p_x,p_y,p_z)$ we have increment $(\Delta p_x,\Delta p_y,\Delta p_z)$ caused by 
some external forces. This increment $(\Delta p_x,\Delta p_y,\Delta p_z)$ has influence on the spin velocity such that 
it appears only in the term $\cos\psi$ in (\ref{9.1}), but not in the torsion and curvature of the trajectory.    

The term $\frac{w\vec{t}\cos \psi}{k}$ leads to the following additional part of the spin velocity 
\begin{equation}
\vec{V}=-\frac{1}{2w^2}\frac{d\vec{b}^*}{dt}(\vec{b}^*\cdot \vec{g}), \label{I}
\end{equation}
if $w$ is sufficiently large, where   
$\frac{d\vec{b}^*}{dt}$ denotes the influence caused by the external forces. 
  
{\bf Example 3.} Let us assume that the unit vector $\vec{b}^*(t)$ of spinning axis of a spinning circle is given by  
$$\vec{b}^*=(\cos \Omega t,0,\sin\Omega t),$$ 
where $\Omega$ is a constant, and assume also that the angular velocity $w$
is much larger constant. According to (\ref{I}) and neglecting the velocities (\ref{9.4}) and (\ref{V-curvature1}),
for the spin velocity we obtain 
\begin{equation}
\vec{V}=\frac{g\Omega \sin\Omega t}{2w^2}(-\sin \Omega t, 0, \cos \Omega t). 
\end{equation}
After averaging by $\alpha$ we see that 
the circle moves with almost a constant velocity 
\begin{equation}
\vec{V}=\bigl (-\frac{g\Omega}{4w^2}, 0, 0\bigr )
\end{equation}
without reaction.    
There appears a periodic acceleration in the 
$z$-axis $A_z=dV_z/dt\neq 0$, but it averages to 0. 

{\bf Example 4.} 

The spin precession of the Earth's axis of about $0.052''/$yr calculated in section 9 and also the precession 
of about $50''$/yr from the torques cause spin velocity 
according to (\ref{9.4}), (\ref{V-curvature1}) and (\ref{I}). In order to estimate the corresponding spin acceleration, 
we should replace $\vec{g}$ by $\vec{g}-\vec{g}_b$, because the Earth is in free  fall motion. Hence it is easy to see 
that the corresponding spin velocity is almost 0, such that these spin velocities do not have any remarkable 
influence on the trajectory of the Earth. 

\section{Conclusion} 
\label{sec:11}

In the first section some anomalies were presented, assuming that the space-time is condensed into 3+1-dimensional space-time. 
Thus starting with the 3+3+3-model which was recently introduced (\cite{Tensor2009,VB,MB}), it is developed now mainly upon the 
structural groups $G_t$ and $G_s$  
and they naturally lead to four structural invariants. Independently from these structural invariants 
a new type of motion - spin velocity was deduced,   
which can be interpreted simply as a displacement in the space.  
Using the deduced spin velocity and precession of the spinning axis, at the end of section 8 
it was shown that the structural invariants are satisfied. 
Although the effects described in this paper are contradictory with Newton's third law in its classical formulation, 
the approach in this paper seems to improve Newton's third law at a more sophistical level, because the considered 
spin motion and spin precession appear to preserve the invariants $J_1,J_2,J_3$ and $J_4$.

\end{document}